\documentclass[preprint,showpacs,preprintnumbers,amsmath,amssymb]{revtex4}
\usepackage{graphicx}
\usepackage{amsfonts} 
\usepackage{amsmath}
\usepackage{amssymb} 

\begin{document}

\title{Monte Carlo simulations on rare-earth Holmium ultra-thin films}

\author{Fabio~Cinti$^{1,2}$}
\email{fabio.cinti@fi.infn.it}
\author{Alessandro~Cuccoli$^{1}$}
\author{Angelo~Rettori$^{1,2}$}
\affiliation{$^{1}$Dipartimento di Fisica, Universit\`a di Firenze, I-50019 Sesto Fiorentino (FI), Italy.}
\affiliation{$^{2}$CNR-INFM S$^3$ National Research Center, I-41100 Modena, Italy.}

\date{\today}

\begin{abstract}
Motivated by recent experimental results in ultra-thin helimagnetic
Holmium films, we have performed an extensive classical Monte Carlo
simulation of films of different thickness, assuming a Hamiltonian 
with six inter-layer exchange constants.
Both magnetic structure and critical properties have been analyzed.
For $n>16$ ($n$ being the number of spin layers in the film) a correct 
bulk limit is reached, while for lower $n$ the film properties are clearly
affected by the strong competition among the helical pitch and the 
surface effects, which involve the majority of the spin layers:
In the thickness range $n=9\div16$ three different magnetic phases emerge,
with the high-temperature, disordered, paramagnetic phase and the 
low-temperature, long-range ordered one separated by an intriguing 
intermediate-temperature block phase, where outer ordered layers coexist 
with some inner, disordered ones. The phase transition of these inner 
layers displays the signatures of a Kosterlitz-Thouless one.
Finally, for $n\lesssim7$ the film collapse once and for all to a
quasi-collinear order. 
A comparison of our Monte Carlo simulation outcomes with available 
experimental data is also proposed, and further experimental
investigations are suggested.

\end{abstract}

%\pacs{65.40.-b,75.40.Cx,75.50.Xx}
%75.40.-s Critical-point effects, specific heats, short-range order (see also 65.40.-b Heat capacities of solids) 
%75.40.Cx Static properties (order parameter, static susceptibility, heat capacities, critical exponents, etc.)  
%75.50.-y Studies of specific magnetic materials
%75.50.Xx Molecular magnets
%75.75.+a Magnetic properties of nanostructures
\maketitle

%%%%%%%%%%%%%%%%%%%%%%%%%%%%%%%%%%%%%%%%%%%%%%%%%%%%%%%%%%%%%%%%%
%%%%%%%%%%%%%%%%%%%%%%%%%%%%%%%%%%%%%%%%%%%%%%%%%%%%%%%%%%%%%%%%%
%%%%%%%%%%%%%%%%%%%%%%%%%%%%%%%%%%%%%%%%%%%%%%%%%%%%%%%%%%%%%%%%%
\section{Introduction}

Surface and nanometrical objects are important both for their 
possible implementation in the current technology and from basic
research point of view.
In collinear magnetic thin films intriguing behavior,
as transition temperature depending from the film thickness $-n-$
or critical exponent crossover, were observed 
\cite{Fisher72,Ritchie,Barber,Henkel99,Henkel98,zhang01} .

Nowadays, the most fervent interest has moved towards
frustrated systems, where a non-collinear order, 
characterized by a possibly large modulation,
is established. Some rare-earth elements
(as Holmium, Dysprosium or Terbium) and their 
compounds are typical examples that the nature makes available
to investigate such peculiar behaviours, in view of the variety 
of magnetic arrangements, as helix, spiral or longitudinal-wave,
that can be observed in bulk samples of such materials.\cite{Jensen91}.
Further examples of helicoidal structures can also be met in 
multiferroic materials\cite{mf1,mf2,mf3}
and itinerant systems, as MnSi \cite{MnSi} and FeGe \cite{FeGe}.

In magnetic systems with frustration, the lack of translational 
invariance due to the presence of surfaces can result especially
important for ultra-thin film samples where the thickness is
comparable, or even lower, with the wave length of the ordered 
magnetic structure observed in the bulk.
It is worthwhile observing that when these conditions are met 
a sweeping change of the magnetic structure behaviour
could be found. 
Many fundamental features related to such systems 
have not yet been exhaustively investigated and completely 
understood,
and ultra-thin films of rare-earth elements 
are still among the most intriguing layered systems to be 
studied\cite{Jensen05}.

From an experimental point of view, the availability of new
sophisticated growth and characterization techniques\cite{Bland}
has allowed to extensively investigate the properties of such 
magnetic nano-structures. For instance, 
interesting experimental data on thin films of 
Holmium (whose bulk samples show helical order along the
$c$ axis, perpendicular to film basal planes) 
were obtained \cite{Weschke04,Weschke05,Jensen05,schss00}
by neutron diffraction and resonant soft X-ray experiments.
By looking at the static structure factor $S(\vec{Q})$ ($\vec{Q}=(0,0,Q_z)$ 
being the wave vector of the incommensurate magnetic modulation 
along the film growth direction $z$), it has been
shown that the critical behaviour of Holmium thin films
markedly differ from that of films of transition metals, 
which usually display a collinear ferromagnetic (FM) 
or antiferromagnetic (AFM) order in the bulk.
Mainly, the authors of 
Refs.~\onlinecite{Weschke04,Weschke05,Jensen05,schss00} identified 
a thickness $n_0\simeq 10$ mono-layers [ML] (comparable with the
helix pitch of bulk Holmium $\simeq 12$\,ML) which was interpreted as 
a lower bound for the presence of the helical ordered phase.
In fact, they observed that the transition temperature 
dependence on film thickness does not follow the usual asymptotic
power law \cite{Barber}, but rather an empirical relation
\begin{equation}
  \label{filmhm}
\frac{T_N(\infty)-T_N(n)}{T_N(n)} \sim
\frac{1}{(n-n_0)^{\lambda^\prime}}~,
\end{equation}
can be devised \cite{Weschke04},
where $T_N(\infty)$ and $T_N(n)$ are the ordering temperature of the bulk
system and of a film with thickness $n$, respectively, 
while the exponent $\lambda^\prime$ has not an universal character.
Interesting enough, it appears that this empirical relation is not
peculiar of helical-like structures only
but it results more general, being observed in other
ultra-thin structures characterized by a magnetic modulation as well:
An important example is given by  Chromium films, where at low 
temperatures an incommensurate spin density wave is
present \cite{Fullerton95}.

A mean field approximation 
(MFA) was proposed in Refs.~\onlinecite{Jensen05} and
\onlinecite{Weschke04} in order to understand the experimental  
outcomes from Holmium films: MFA allowed to obtain a first,
rough estimate of the threshold thickness $n_0$ defined in 
the empirical relation~\eqref{filmhm}, but it also revealed 
that for thicknesses close enough to $n_0$ the paramagnetic 
and helical phase can be accompanied by a more complex block 
phase, where groups of ferromagnetically ordered layers 
pile-up in an alternating antiferromagnetic arrangement 
along the c axis. As it is well known, the MFA completely 
neglects thermal fluctuations, which are however strongly expected to 
play a fundamental role in the critical behaviour 
of low-dimensional magnetic systems: therefore, adding thermal 
fluctuations not only give strong quantitative adjustments
of the MFA estimates of critical quantities as $T_N(n)$ or $n_0$,
but could also make unstable some ordered structure as the block phase
met above.

In order to overcome such issues and deepen our understanding 
of critical phenomena in Holmium ultra-thin films,
we performed extensive classical Monte Carlo
Simulations (MCS): Preliminary results already showed \cite{Cinti1}
that thermal fluctuations do not destroy the block phase,
which instead acquire a much richer structure, with 
disordered inner layers intercalating ordered ones and 
undergoing a Kosterlitz-Thouless (KT) phase transition\cite{KTt} 
as the temperature lowers.

A complete account is here given of the results of our simulations 
for film thickness in the range $n=6\ldots36$\,ML.
The paper is organized as follows: In Sec.~\ref{sec1} we shall
briefly recall relevant properties of Holmium, and 
introduce the magnetic model Hamiltonian. Sec.~\ref{mcm} is devoted
to describe the Monte Carlo method and the estimators employed to 
evaluate the physical quantities relevant for magnetic films with 
non-collinear order.
In Sec.~\ref{lowt} the Monte Carlo results about the magnetic 
order establishing at low temperature are shown for different 
thicknesses.
The role of thermal fluctuations is discussed in Sec.~\ref{fluctu}:
In particular, Sec.~\ref{planes} will report 
a detailed study of the temperature regions where 
the single layers display a critical behaviour, the structure 
factor close to these regions being deeply analyzed too,
given its fundamental relevance in an experimental mindset;
Sec.~\ref{gfp} will be devoted to the global film properties.
All the results reported in the previous sections will be 
compared and discussed in an unifying framework in Sec.~\ref{disc},
where we shall also gather our conclusions.

%%%%%%%%%%%%%%%%%%%%%%%%%%%%%%%%%%%%%%%%%%%%%%%%%%%%%%%%%%%%%%%%%
%%%%%%%%%%%%%%%%%%%%%%%%%%%%%%%%%%%%%%%%%%%%%%%%%%%%%%%%%%%%%%%%%
%%%%%%%%%%%%%%%%%%%%%%%%%%%%%%%%%%%%%%%%%%%%%%%%%%%%%%%%%%%%%%%%%
\section{Model Hamiltonian}\label{sec1}

The magnetic properties of Holmium have been intensively 
investigated both experimentally and theoretically \cite{Jensen91}. 
The bulk crystal structures is known to be hexagonal close-packed 
(\textit{hcp}). The indirect exchange among the localized 
\textit{4f} electrons manifests as an RKKY long-range interaction 
of atomic magnetic moments; the experimental data about the 
low-temperature magnetic moments arrangement in 
Ho can be reproduced assuming a FM interaction between nearest 
neighbor spins lying on the \textit{ab} crystallographic 
planes\cite{Jensen91}, while along the \textit{c}
crystallographic axis interactions up to the sixth neighboring 
layers must be allowed for (see for example Ref.~\onlinecite{Borh89}).
It's just the competing nature of the latter that below $T_N(\infty)=132$\,K
gives rise to an incommensurate magnetic periodic structure,
which can be modeled as an helical arrangement of the magnetic moment 
vectors along the direction (henceforth denoted as \textit{z}) parallel 
to \textit{c}, i.e. perpendicular to the \textit{ab} crystallographic planes, 
where the magnetic vectors prefer to lye as a consequence of a 
single-ion easy-plane anisotropy.
The average local spin vector at low temperature can thus be expressed as:
\begin{equation}
\label{sqord}
\vec{s}_i\equiv\vec{s}\left(\vec{r}_i\right) = 
\left(s_{\perp}\cos\left(\vec{Q}\cdot\vec{r}_i\right),
s_{\perp}\sin\left(\vec{Q}\cdot\vec{r}_i\right),s_z\right)~,
\end{equation}
where \(\vec{Q}=\left(0,0,Q_z^{bulk}\right)\) is the helical pitch vector, and
$c\,Q_z^{bulk}\simeq \pi/3$, \textit{c} being the lattice constant \cite{Jensen91}
along the \textit{z} direction (i.e., $c/2$ is the distance between 
nearest neighbouring \textit{ab} spin layers). In addition, the crystal field 
bring into play other different kinds of anisotropies that, at temperatures
well below $T_N(\infty)$, are able to change the helical shapes in conical ones, 
or force the magnetic structures in a \textit{bunched helix} which is 
commensurate with the lattice \cite{Gibbs85,Jensen87}.

Theoretical investigations have shown that incommensurate magnetic bulk 
structures (observed, besides Holmium, also in other rare-earth elements, 
as Dysprosium and Terbium) can be well obtained by a
MFA \cite{Coqblin77} from a simple Heisenberg model with only 
three coupling constant, the first one, $J_0>0$, describing the FM in-plane 
interactions, while $J_1$ and $J_2$ are the effective coupling between 
ions on neighbouring (NN) and next-neighbouring (NNN) planes, respectively. 
Whatever the sign of $J_1$, the MFA finds a helical structure when 
$J_2<0$, i.e. AFM, and the conditions $|J_2|>|J_1|/4$ is met. 

\begin{figure}
  \begin{center}
    \includegraphics[width=0.6\textwidth]{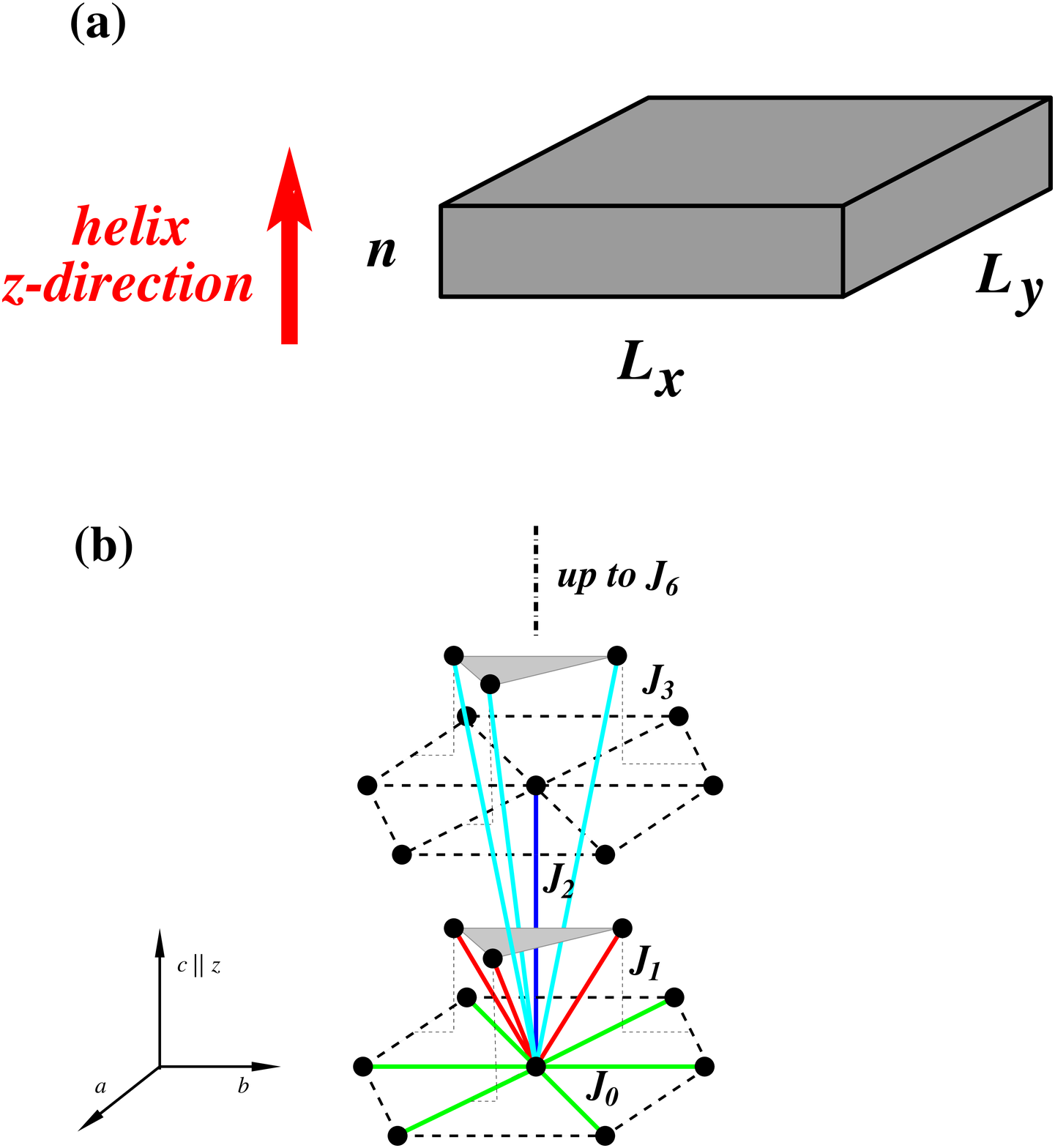}
  \end{center}
  \caption[]{(color online) 
    \textbf{(a)} Film geometry: $N=n\times L_x \times L_y$ is the total number 
    of spins, $n$ is the film thickness, i.e. the number of spin layer 
    (free boundary conditions are taken
    along $z$-direction), and $L_{x,y}$ are the layer dimensions 
    (periodic boundary conditions are applied along $x$ and $y$
    directions).
    \textbf{(a)} Schematic representation of 
    \textit{hcp} Ho structure and exchange interactions: only $J_0$ 
    (in-plane, green lines), $J_1$ (NN planes, red lines), $J_2$ 
    (NNN planes, blue line), and $J_3$ (cyan lines) are represented. 
    The numerical values of the coupling constants employed in our 
    simulations are: $J_0=55.68$\,K, $J_1=24.9168$\,K, $J_2=5.4288$\,K, 
    $J_3=-3.3686$\,K, $J_4=-4.5936$\,K,  $J_5=-0.2784$\,K, $J_6=-2.5056$\,K.}
  \label{hcplatt}
\end{figure}

It is worthwhile to recall that when dealing with ultra-thin films
the assumption of being allowed to retain the same Hamiltonian able 
to describe the bulk structure is absolutely not guaranteed 
to be correct: Indeed, real film samples can be strongly 
affected by defects, strain, thickness uncertainty ($\sim$2 ML) or 
interaction with the substrate \cite{comm} (typically Y/Nb or W(110)). 
The latter can be particularly relevant, as it can change, sometimes 
dramatically, the single ion anisotropy and the 
strength of the interaction constants with respect to bulk samples.
Furthermore, the lack of inversion symmetry can bring into play 
the antisymmetric Dzyaloshinskii-Moriya (DM) interaction
\cite{moriya60} (see for instance Ref.~\onlinecite{Dagotto06} and
\onlinecite{Mostovoy06} for perovskite multiferroic RMnO$_3$ with
R\,=\,Gd,\,Tb, or Dy) and possible surface anisotropies for Dy/Y
multilayer films \cite{super}.

While always remembering such possible drawbacks, in our investigation 
of Ho thin films we employ a Heisenberg model Hamiltonian which has 
proven useful to describe Holmium bulk samples. We thus define:
\begin{equation}
\label{Hamiltonian}
{\cal H} = -\sum_{i,j}^{N} J_{ij} \vec{\sigma}_i \cdot \vec{\sigma}_j + 
D_z\sum_{k=1}^{N} \left(\sigma^{z}_{k}\right)^{2}~,
\end{equation}  
where $N=n\times L_x \times L_y$ is the total number of magnetic ions, 
$L_x=L_y=L$ being the lateral film dimensions and $n$ the number of 
layers (see Fig.~\ref{hcplatt}a), and
$\vec{\sigma}_i$ are classical unitary vectors representing the total 
angular momentum of Ho ions (i.e. $\vec{\sigma}_i = 
\vec{\mathcal{J}}_i/|\vec{\mathcal{J}}_i|$). As a consequence, the magnitude
of the angular momentum of Ho ions is embodied in our definition of 
the coupling constants appearing in Eq.~\eqref{Hamiltonian} as 
$J_{ij} = \frac{1}{2}|\vec{\mathcal{J}}|(|\vec{\mathcal{J}}|+1)\textsc{J}_{ij}$, 
with $|\vec{\mathcal{J}}_i|=8$. For the inter-layer exchange parameters 
$\textsc{J}_{ij}$ we have assumed the values given in 
Ref.~\onlinecite{Borh89}, while the easy plane
anisotropy $D_z=16$\,K \cite{Coqblin77}.

In Fig.~\ref{hcplatt}b we show a schematic representation of 
the \textit{hcp} lattice structure and of the exchange interactions 
included in the model Hamiltonian~\eqref{Hamiltonian}. On the hexagonal 
basal planes only a NN, FM interaction $J_0>0$ (green lines) is considered;
along the $c$-axis we instead allow for interactions up to the sixth 
neighboring layer, with a total coordination number $\zeta=30$.

Sign and magnitude of the out-of-plane interaction constants 
J$_1$ ... J$_6$ were determined in Ref.~\onlinecite{Borh89} 
in order to reproduce, with the correct pitch vector, the 
helical ground state along the \textit{c}-axis observed in 
the experimental data.
However, at the best of our knowledge, also neutron scattering 
experiments investigated the dynamical properties of Holmium
only along the $c$-axis \cite{Jensen87}, so that a direct measure 
of the in plane FM coupling constant is still lacking, and only
mean field estimates are available, which set it at about 
300$\mu$eV \cite{Weschke04}. This allows us to consider J$_0$ as an 
almost free fit parameter to be adjusted in order to fix the correct 
value  of experimentally accessible quantities. By Monte Carlo 
simulations we find the value 
$T_N^{\textnormal{MCS}}(\infty) \simeq 124$\,K for the bulk ordering 
temperature by setting J$_0\approx 400 $\,$\mu$eV: bearing in mind 
the cautions given above about the possible quantitative difference 
among bulk and film samples, this is the value of J$_0$ we have used 
in all the following simulations of thin films, without attempting 
any further possibly meaningless quantitative adjustment.

%%%%%%%%%%%%%%%%%%%%%%%%%%%%%%%%%%%%%%%%%%%%%%%%%%%%%%%%%%%%%%%%%
%%%%%%%%%%%%%%%%%%%%%%%%%%%%%%%%%%%%%%%%%%%%%%%%%%%%%%%%%%%%%%%%%
%%%%%%%%%%%%%%%%%%%%%%%%%%%%%%%%%%%%%%%%%%%%%%%%%%%%%%%%%%%%%%%%%
\section{Monte Carlo Methods and Thermodynamic Observables}\label{mcm} 
Our study of the magnetic properties of thin rare earth films was 
done by extensive classical MCS.
Thickness \textit{n} from 6 to 36 and lateral dimensions 
$L_x=L_y=L=8\ldots80$ have been analyzed. As we are working on 
film structures, free boundary conditions in the thickness direction 
$z$ are obviously taken, while the usual periodic boundary conditions
are applied in the $L\times L$ planes (see Fig.~\ref{hcplatt}a), which 
coincide both with the \textit{ab} crystallographic planes and the easy 
plane for the magnetization.

Simulations were done at different temperatures; the thermodynamic 
equilibrium is reached by the usual Metropolis algorithm \cite{met53} 
and over-relaxed technique \cite{over87}. The latter was employed in order 
to speed-up the sampling of the whole spin configuration space:
Indeed, the competitive nature of the exchange interactions 
in the Hamiltonian~\eqref{Hamiltonian} and the high coordination number
lead to a long time needed to reach thermal equilibrium, especially in 
the critical region.
We have thus resorted to a judicious mix of Metropolis and over-relaxed 
moves in order to reach the goal in a reasonable time. Usually, one 
``Monte Carlo step'' is composed by one Metropolis and four/five 
over-relaxed moves per particle, discarding up to 5$\times$10$^4$ 
Monte Carlo steps for thermal equilibration; at least three independent
simulations where done for for each temperature.

The Monte Carlo data analysis have benefited from the employment of 
the multiple histogram technique \cite{Ferrenberg88,bootstrap}, 
which allows us to estimate the physical observables of interest over 
a whole temperature range by interpolating the results obtained from 
simulations performed at some chosen, different temperatures. 
The outcome of the method is an estimate of the density of state 
$\rho(\mathcal{E})$ at energy $\mathcal{E}$ obtained by weighting 
the contributes $\rho_i(\mathcal{E})$ due to independent simulations 
made at inverse temperature $\beta_i$ ($\beta_i=1/k_BT_i$). The 
independent simulations have to be sufficiently close in temperature, 
i.e. the temperature step must be chosen roughly proportional to the
square root of the inverse of the heat capacity \cite{bootstrap}. 
We thus have computed the partition function $\mathcal{Z}_\beta$ at any 
$\beta$ in the range of interest by solving iteratively the equation
\begin{equation}
\label{zeta}
\mathcal{Z}_\beta = \sum_{\mathcal{E}}\rho(\mathcal{E})\textnormal{e}^{-\beta\mathcal{E}}=
\sum_{i,s}\left[\sum_k\frac{m_k}{\mathcal{Z}_{\beta_k}}\textnormal{e}^{\Delta\beta_k\mathcal{E}_{i,s}}\right]^{-1} ~,
\end{equation}  
where $\Delta\beta_k=\beta-\beta_k$ and $m_k$ is the number of 
independent samples of energy for the $k$-simulation. 
The index $i$ refers again to the single simulation, while $s$ denotes 
the energy sampling intervals in the $i$-th simulation.

As mentioned above, the high number of exchange interactions 
makes difficult the estimate of the density of state, especially 
close to the critical temperature, and variables like specific 
heat or susceptibility are extremely sensitive for these systems. 
Anyway, this obstacle has been successfully overcome making 
use of the estimator,
\begin{equation}
\label{cv}
C_v = \frac{\beta^2}{N}\langle\left(\mathcal{E}-\langle \mathcal{E}\rangle_\beta\right)^2\rangle_\beta
\end{equation}  
in the histogram reweighting techniques, as suggested in 
Ref.\onlinecite{PRElandau}. 
Iterating several times the multiple-histogram algorithm, we have also
obtained the variance of the interpolated data by bootstrap resampled 
method, by picking out randomly a sizable number of 
independent measurements $m_k$ (between 1 and 5$\times$10$^{4}$), and iterating the 
re-sampling at least one hundred times \cite{bootstrap,bootstrap1}.

As we are interested in the phase transitions of Holmium films,
it is worthwhile to observe that the study of films described by the 
Hamiltonian~\eqref{Hamiltonian} entails a wide number of fundamental 
issues. First of all we must consider: \textit{i)} the intrinsically 
two-dimensional (2d) nature of such magnetic structures; \textit{ii)}
the presence of different interactions,
which turn out to be FM on the layers 
(with a SO(2) symmetry) and competitive 
along the thickness direction \textit{n}, with a possible helimagnetic (HM) 
order at low temperature (i.e. a $\mathbb{Z}_2$ $\times$SO(2) 
symmetry \cite{Kawamura});
and \textit{iii)} the implementation of different boundary conditions 
for \textit{n} and \textit{L}, respectively, above introduced.

About the first issue, it is well known that the critical behaviour of 
an ideal easy-plane magnetic film with continuous symmetry and short-range FM 
(or AFM) interactions pertains to the 2d XY universality class,
displaying a KT behaviour \cite{KTt} at a finite 
critical temperature. In particular, a crossover from 3d to 2d behaviour 
is expected when the correlation length saturates the film thickness.
However, from MCS point of view, it may result 
quite difficult to realize such conditions. Indeed, even for large, but 
still finite, $L$ values a sharp transition can not be observed, making 
it possible to define a \textit{3d pseudo critical point}, as
extensively discussed by Janke and co-workers in Refs.~\onlinecite{Janke93a,Janke93b}.

Turning to the second issue, in a quasi-2d magnetic system with a continuous 
symmetry the introduction of competing interactions along the direction 
perpendicular to the film slab brings into play the presence of two 
(in-plane and out-of-plane) correlation lengths, 
with a rather dissimilar behaviour in the critical regime. As analyzed 
in Ref.~\onlinecite{Cinti1}, under these conditions some new and interesting 
critical phenomena can be observed. 
Systems with discrete
symmetries which present two different correlation lengths were
already discussed in literature, see, e.g., 
Refs.~\onlinecite{wang89} and \onlinecite{bunker93}.

Moving to the last issue, we must first of all observe that 
the identification of a suitable order parameter to study the critical 
properties of non-collinear thin films requires a careful analysis of
some of their peculiar features.
A first trouble is the intrinsic difficulty represented by a 
helical order parameter associated to a wave vector $\vec{Q}$:
In the bulk system the virtually infinite size of the system, 
summing up an \textit{infinite} number of in-phase contributions,
makes a clear peak emerge at wave-vector $\vec{Q}$ 
in the static structure factor in the ordered phase.
In films, the presence of broad peaks is on the contrary expected in a wide 
temperature range as, a consequence of the intrinsic finite size nature of 
the system, thus jeopardizing the identification of a well defined critical 
temperature $T_N(n)$ from the sole analysis of the peaks appearing in the structure 
factor. Secondly, as we will discuss in the next Sections, a \textit{na\"ive} structure factor analysis 
could be not enough to distinguish between an HM order and other 
ordered phases that can be present.
For these reasons, it is necessary to resort to other observables related to 
the HM order. A first choice can be found in the chirality \cite{Diep94,Cinti2}, 
which can be defined on film as:
\begin{equation}
\label{Chirality}
\kappa = \frac{1}{3(n-1)L^2\sin Q_z^{bulk}}\sum_{li}\left(\vec{\sigma}_{l,i}
\times\vec{\sigma}_{l+1,i}\right)^{z}~,
\end{equation}   
where $l$ labels the planes, starting from one of the two film surfaces, 
and $i$ locates the spin on the plane.
As we shall see in the next Sections $\kappa$ represents quite 
a good quantity to locate the critical temperature for the HM phase. 
In view of point \textit{i)} discussed above, it is useful to introduce a FM 
order parameter for each layer $l$:
\begin{equation}
\label{mplane}
M_l = \sqrt{(M^x_l)^2 + (M^y_l)^2 + (M^z_l)^2}~,
\end{equation} 
where $M^\alpha_l = \frac{1}{L^2}\sum_i \sigma^\alpha_{l,i}$, with $\alpha=x,y,z$,
and consequently the average order parameter of the film \cite{Loison00a,Loison00b}:
\begin{equation}
\label{mave}
M = \frac{1}{n}\sum_l^n M_l ~.
\end{equation}   
It is worth to observe that the order signalled by the quantities defined in 
Eq.~\eqref{mplane} and \eqref{mave} do not directly entail the existence of 
an HM or fan structure in the film.

The critical nature of the observables defined in Eqs. \eqref{Chirality}, 
\eqref{mplane}, and \eqref{mave}, is better revealed by looking at the 
following derived quantities ($\mathcal{O}=\kappa,M_l,M$):
\begin{equation}
\label{susce}
\langle\chi_{\mathcal{O}}\rangle=N\beta\left(\langle\mathcal{O}^2\rangle
-\langle\mathcal{O}\rangle^2\right) ~,
\end{equation}   
\begin{equation}
\label{opprime}
\frac{\partial}{\partial\beta}\langle\mathcal{O}\rangle=
\langle\mathcal{OE}\rangle-\langle\mathcal{O}\rangle\langle\mathcal{E}\rangle ~,
\end{equation}  
\begin{equation}
\label{lnopprime}
\frac{\partial}{\partial\beta}\langle\ln\mathcal{O}\rangle=
\frac{\langle\mathcal{OE}\rangle}{\langle\mathcal{E}\rangle}
-\langle\mathcal{O}\rangle\langle\mathcal{E}\rangle~,
\end{equation}  
which, at the critical temperature, display a peak that can be 
characterized by the usual finite size scaling theory \cite{Libbinder}.
In particular, for large enough $L$, $T_N^L(n)$ approximately scales as 
\begin{equation}
\label{tscaling}
T_N^L(n) \approx T_N^\infty(n)+CL^{-1/\nu}  ~,
\end{equation}  
where $C$ is system dependent constant, while $\nu$ is the 
correlation length critical exponent.

A final quantity we employ in our investigation of the critical 
properties of films is the Binder cumulant \cite{binderu4,binderprl}:
\begin{equation}
\label{bcum1}
U_4 =  1 - \frac{\langle\mathcal{O}^4\rangle}{\langle\mathcal{O}^2\rangle^2}~,
\end{equation}   
which allows us to locate the critical temperature by looking at the 
intersection of the graphs of $U$ as a function of $T$ obtained 
at different $L$: At $T=T_{N}(n)$ such crossing becomes a ``nontrivial
fixed-point'' \cite{binderu4,binderprl}. Moreover,
one can examine the ratios $U_{L^\prime}/U_{L}$ (for sizes $L$ and
$L'>L$) as temperature function, looking for a unitary ratio
at the critical temperature.

%%%%%%%%%%%%%%%%%%%%%%%%%%%%%%%%%%%%%%%%%%%%%%%%%%%%%%%%%%%%%%%%%
%%%%%%%%%%%%%%%%%%%%%%%%%%%%%%%%%%%%%%%%%%%%%%%%%%%%%%%%%%%%%%%%%
%%%%%%%%%%%%%%%%%%%%%%%%%%%%%%%%%%%%%%%%%%%%%%%%%%%%%%%%%%%%%%%%%
\section{Magnetic Structures at low Temperature}\label{lowt}

In this section we will present and analyze our Monte Carlo results for 
the overall magnetic behaviour of film samples of different thicknesses, 
from a bulk-like structure with $n=36$ to a very thin film of $n=6$ layers,
at temperature $T=10$\,K, i.e. well below $T_N(n)$; 
the lateral dimension of the films is taken constant at $L=80$,
having checked that at this temperature, far from the critical region,
such value of $L$ well represents the thermodynamic 
limit for all practical purpose.

\begin{figure}
  \begin{center}
    \includegraphics[width=0.8\textwidth]{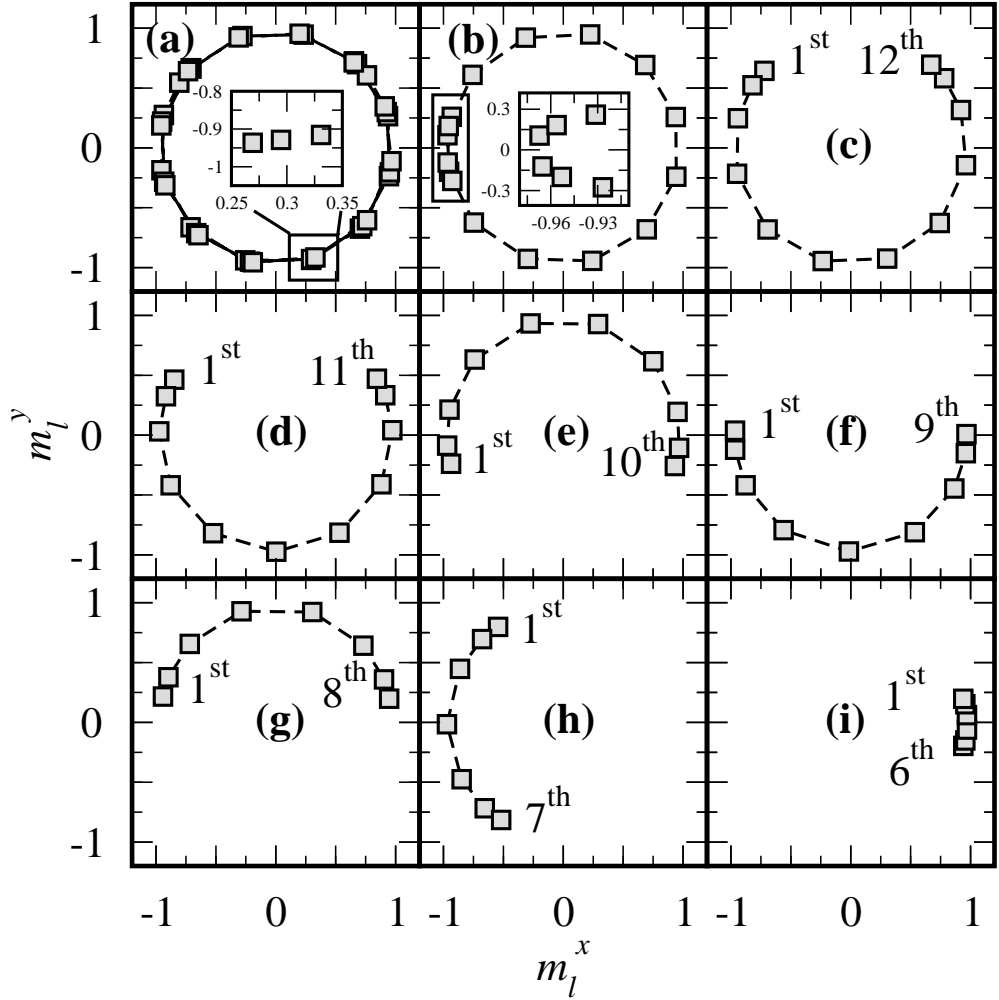}
  \end{center}
  \caption[]{Normalized magnetic vector profiles of each layer 
    at $T=10$\,K, layer dimensions $L_{x,y}=80$ and different 
    film thickness $n$: 
    \textbf{(a)} $n=36$; \textbf{(b)} $n=16$; \textbf{(c)} $n=12$; 
    \textbf{(d)} $n=11$; \textbf{(e)} $n=10$; \textbf{(f)} $n=9$; 
    \textbf{(g)} $n=8$; \textbf{(h)} $n=7$; \textbf{(i)} $n=6$;
    the error bars are included in point dimensions. Inset 
    in \textbf{(a)}: the magnified zone of the main graph (here shown for
    the 6$^{\textnormal{th}}$, 18$^{\textnormal{th}}$, and
    30$^{\textnormal{th}}$ planes from left to right along
    $m_l^x$, respectively) clearly
    shows that the helix pitch and lattice structure are incommensurate;
    inset in \textbf{(b)}: the magnification shows the magnetization 
    of the first surface layers: 1$^{\textnormal{st}}$,
    14$^{\textnormal{th}}$, 2$^{\textnormal{nd}}$, 15$^{\textnormal{th}}$,
    3$^{\textnormal{rd}}$, and 16$^{\textnormal{th}}$ from top to bottom along the $m_l^y$, respectively.}
  \label{mxy}
\end{figure}
In Fig.~\ref{mxy} the normalized in-plane magnetic vector 
$\left(m_{l}^x,m_{l}^y\right)$ profile of each layer $l$ 
is reported: For thicknesses greater than 12\,ML 
(which coincide roughly with the helix pitch of bulk Holmium) a behavior 
essentially unaffected by surface effects is observed in
almost the whole sample, with a typical HM order 
(Fig.~\ref{mxy}a-b, i.e. $n=36-16$, respectively). 
 
\begin{figure}
  \begin{center}
    \includegraphics[width=0.8\textwidth]{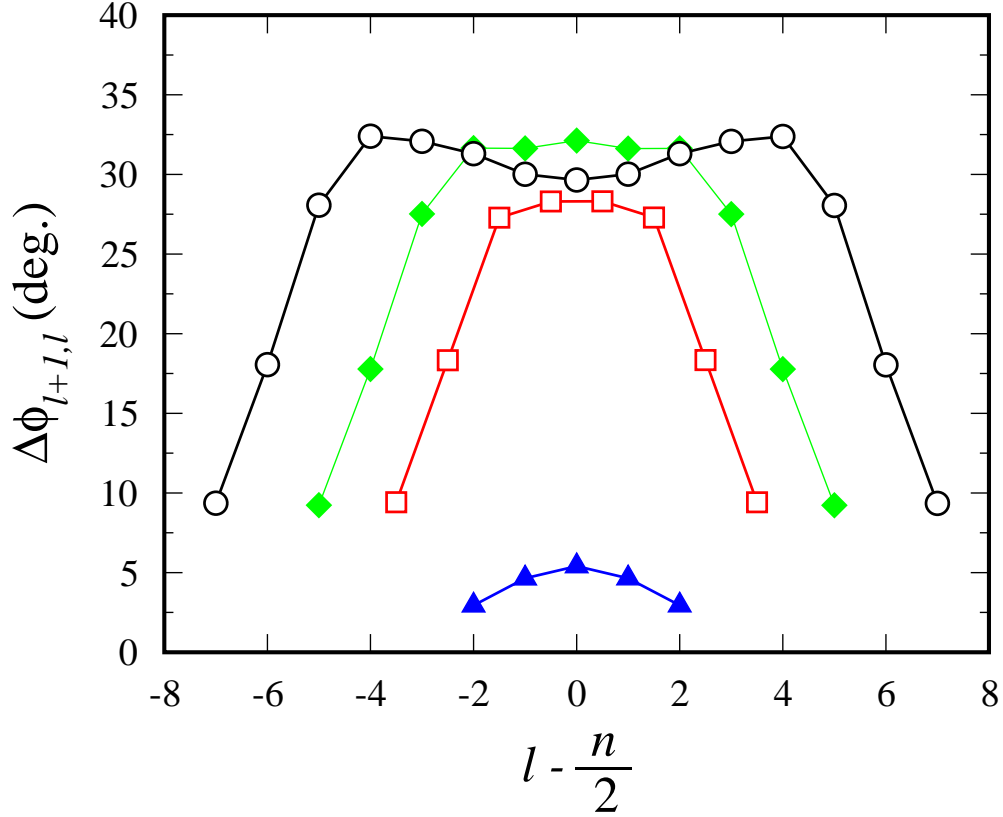}
  \end{center}
  \caption[]{(Colors online) Angle $\Delta\phi_l=\phi_{l+1}-\phi_l$ 
    between magnetic moments in NN layers $(l+1,l)$ at $T=10$\,K, 
    and $L=80$, for representative thicknesses: $n=16$ 
    (black circle), $n=12$ (green diamond), $n=9$ 
    (red square), and $n=6$ (blue up-triangle). 
    Error bars included in the point dimensions.}
  \label{dphi}
\end{figure}
As emphasized in the the inset of Fig.~\ref{mxy}b the 
magnetizations of NN planes close to the surfaces form angles well 
lower than those observed in the bulk, an expected consequence of 
the increasing lack of interactions on one side of the planes 
as the surface is approached. Such effect can be better looked at by 
defining the magnetization rotation angle between NN planes 
$\Delta\phi_l\equiv\phi_{l+1}-\phi_l$.
In Fig.~\ref{dphi} $\Delta \phi_l$ is 
displayed for some representative values of the thickness: 
for thick samples, surface effects are especially strong only on 
the first three layers on each film side, and this explains 
why while for $n\gtrsim9$ an almost bulk behaviour can be 
observed, at least for some inner planes, the scenario changes 
significantly when $n$ drops below 9 (Fig.~\ref{mxy}f-h).

\begin{figure}
  \begin{center}
    \includegraphics[width=0.8\textwidth]{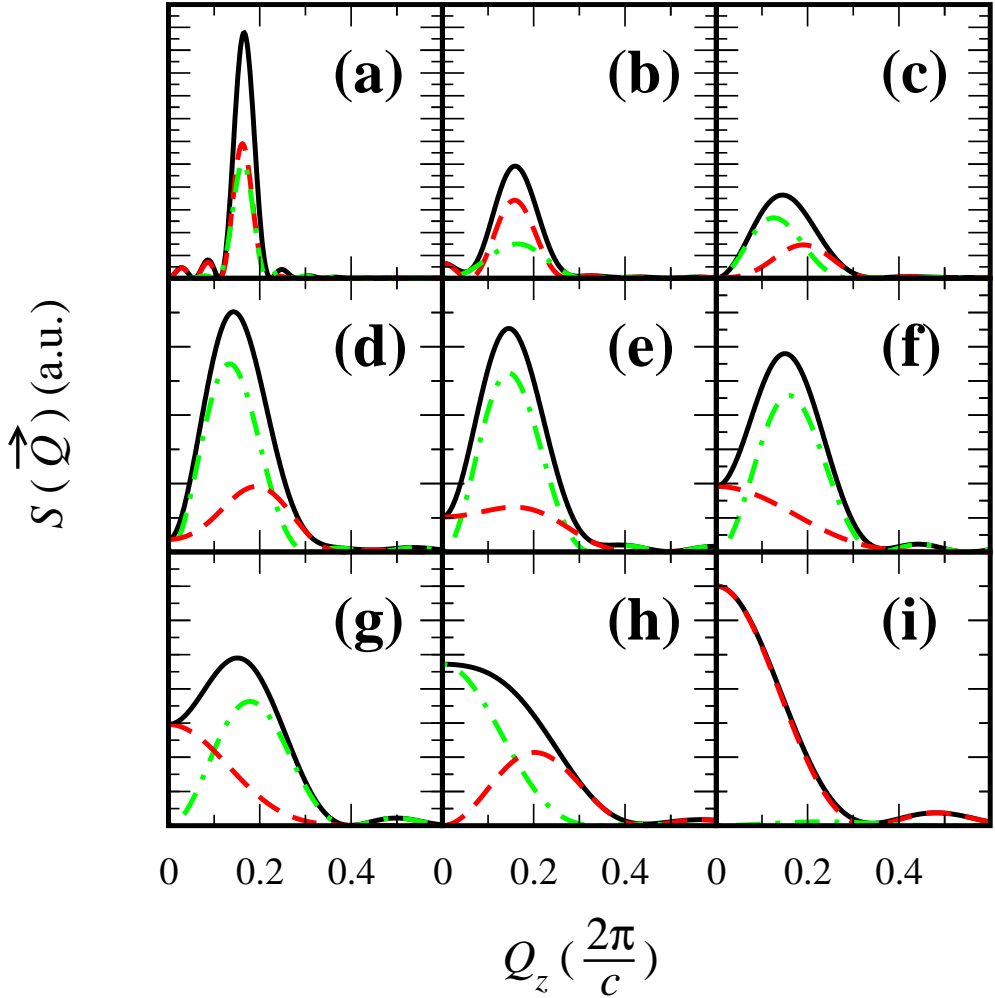}
  \end{center}
  \caption[]{(Colors online) Structure factor $S(0,0,Q_z)$ (continuous line) \textit{vs.} $Q_z$ 
    at $T=10$\,K and layer dimensions $L_{x,y}=80$ for different film 
    thicknesses: \textbf{(a)} $n=36$; \textbf{(b)} $n=16$; 
    \textbf{(c)} $n=12$; \textbf{(d)} $n=11$; \textbf{(e)} $n=10$; 
    \textbf{(f)} $n=9$; \textbf{(g)} $n=8$; \textbf{(h)} $n=7$; 
    \textbf{(i)} $n=6$. Red dashed lines and green dot-dashed lines are the structure factor
    components along the $x$ or $y$ spin space directions, see text. $Q_z$ is measured
    in reciprocal lattice units $2\pi/c$}.
  \label{sqtot0}
\end{figure}
The characterization of the magnetic order can be further pursued
by looking at the static structure factor $S(\vec{Q})$ 
(where $\vec{Q}=(0,0,Q_z)$), i.e. to the 
Fourier transform of the spin correlation function along the $z$-direction 
of the films. $S(\vec{Q})$ is reported in Fig.~\ref{sqtot0} 
(continuous line), together with its in-plane components $S^{xx}(\vec{Q})$ 
and $S^{yy}(\vec{Q})$ (red and green line, respectively; $x$ and $y$ 
directions do not obviously have any special meaning and can be chosen 
at will: here we use the same orientation already employed in 
Fig.~\ref{mxy}).
Once again, for $n\gtrsim11$ (Figs.~\ref{sqtot0}a-d) both the global
structure factor and its components shows a clear peak at
$Q_z^{max}\simeq 0.17$, a value in total agreement with the bulk one
$Q_z^{bulk}\simeq1/6$.
On the contrary, for $n\leqslant 10$ a fan-like structure appears, 
signalled by the emergence of an FM component (i.e. a maximum at $Q_z=0$)
in  $S^{xx}(\vec{Q})$ or $S^{yy}(\vec{Q})$ (Fig.~\ref{sqtot0}f-h), while for 
$n\lesssim7$ a quasi-collinear spin arrangement is finally reached, 
as testified by the single maximum of $S(\vec{Q})$ itself at 
$Q_z^{max}=0$ (Fig.~\ref{sqtot0}i).

The results discussed so far show that, at low temperature, the progressive film thickness reduction 
does not seem to lead to a sudden helical order suppression, but rather 
to induce a gradual passage to a fan-like order associated with a helix
distortion due to the surface effects, until a permanent collapse to
an almost collinear order occurs for $n\lesssim 7$. 
Summing up we can roughly assume that MCS data show that for thickness 
$n<9$ the helical order is substantially absent.

Such results can be considered in fairly agreement with the experimental 
outcomes: In fact, in Ref.~\onlinecite{Weschke04}
the authors identified the thickness $n_0\simeq 10$\,ML
as the value indicating the complete lack of helical order,
and a thickness uncertainty around about 2\,ML\cite{comm} must be taken 
into account.

%%%%%%%%%%%%%%%%%%%%%%%%%%%%%%%%%%%%%%%%%%%%%%%%%%%%%%%%%%%%%%%%%
%%%%%%%%%%%%%%%%%%%%%%%%%%%%%%%%%%%%%%%%%%%%%%%%%%%%%%%%%%%%%%%%%
%%%%%%%%%%%%%%%%%%%%%%%%%%%%%%%%%%%%%%%%%%%%%%%%%%%%%%%%%%%%%%%%%
\section{Magnetic structures in the order-disorder boundary region}\label{fluctu}

\subsection{Layer's Magnetic Behaviour}\label{planes}

This subsection is devoted to the investigation of single layer critical 
properties for some $n$ values.
At the beginning, our attention will be focused on thicknesses close to the 
Ho helix pitch (12\,ML).
For this purpose, the order parameter for each spin layer, as defined in 
Eq.~\eqref{mplane}, is evaluated, together with its Binder cumulant, 
Eq.~\eqref{bcum1}, and its susceptibility,Eq.~\eqref{susce}.
Hereafter, we will denote with the symbol $T_{C,n}$($l$) the transition 
temperature of the $l-$th layer of the film of thickness $n$.

\begin{figure}
  \begin{center}
    \includegraphics[width=1.0\textwidth]{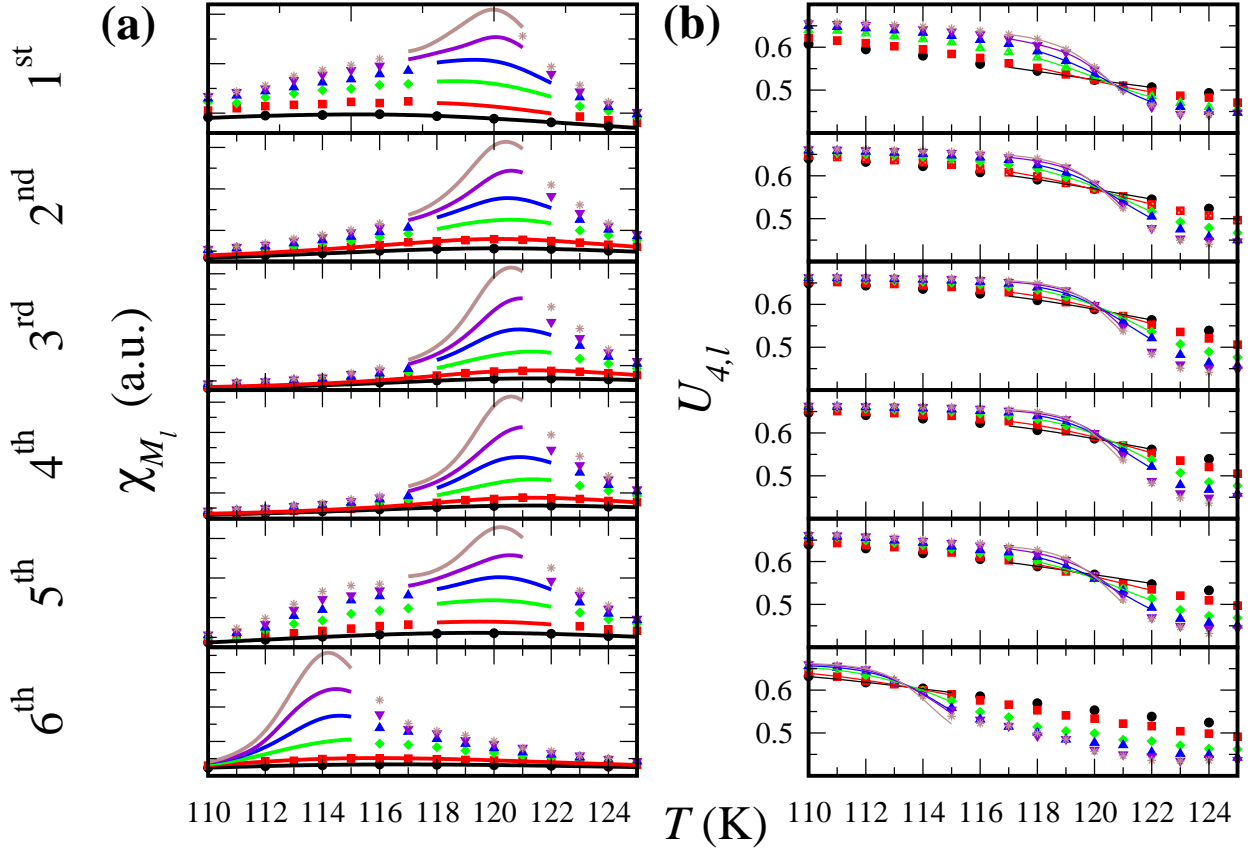}
  \end{center}
  \caption[]{(color online) Susceptibility \textbf{(a)}, and Binder cumulant 
    \textbf{(b)} \textit{vs.} temperature, of the order parameter $M_l$ 
    for thickness $n=12$ and $L= 12$ (black circle), 16 (red square), 24 
    (green diamond), 32 (blue up-triangle), 40 (purple down-triangle), 
    and 48 (brown star). $l$ runs from the first to the sixth spin layer. 
    Close to the critical temperature $T_{C,12}$($l$), 
    the continuum lines are obtained by multiple histogram technique.}
  \label{pippo}
\end{figure}
Susceptibility and Binder cumulant for the first six layers are 
plotted in Fig.~\ref{pippo} as a function of temperature for different 
values of lateral dimension $L$ at $n=12$.
A critical region in a wide temperature range around $T\sim120$\,K is
observed in Fig.~\ref{pippo}a for all planes but the central ones 
(the 6$^{\textnormal{th}}$ and, for symmetry reason, the 
7$^{\textnormal{th}}$) which are definitely still in a paramagnetic 
state, displaying instead a critical region shifted around $T\sim 114$\,K. 
Using the Binder cumulant, Eq.~\eqref{bcum1}, for different values of $L$ we can estimate the
single layer transition temperature $T_{C,12}(6)=113.4(4)$\,K of the inner planes 
and $T_{C,12}(1\ldots5)=120.3(4)$\,K of the external ones. 
\begin{figure}
  \begin{center}
    \includegraphics[width=0.5\textwidth]{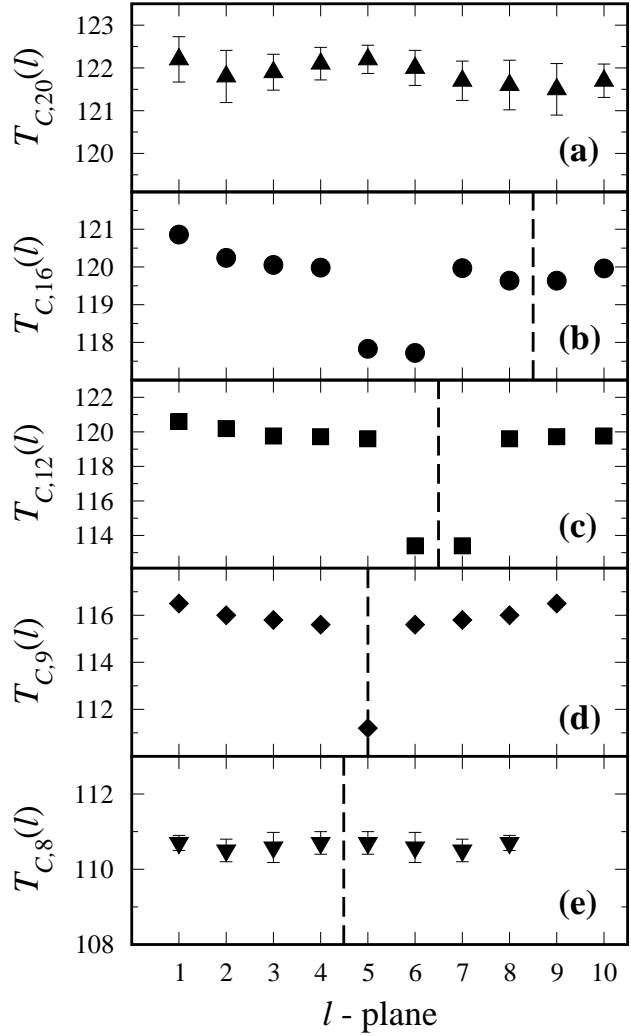}
  \end{center}
  \caption[]{$T_{C,n}$($l$) \textit{vs.} layer index $l$ for $n=20$ \textbf{(a)}, 
    16 \textbf{(b)}, 12 \textbf{(c)}, 9\textbf{(d)}, and 8\textbf{(e)}. 
    The vertical dashed lines denote the position of the bisecting plane 
    of the film, beyond which single-layer properties repeat by symmetry.}
  \label{pippo1}
\end{figure}

The intriguing landscape here observed for $n=12$
is present in the whole range $9\le n\lesssim 16$.
A summarizing picture of the single layer critical temperature 
$T_{C,n}$($l$) \textit{vs.} plane index $l$ for $n=$\,20,\,16,\,12,\,9 and 8 
is given in Fig.~\ref{pippo1}:
For the thicker film here analyzed ($n=20$, Fig.~\ref{pippo1}a)
$T_{C,20}$($l$) is the same for every layer, and coincides with the establishing of
HM order in the film, as expected for the bulk system, where the 
critical temperature can be obtained both through the 
chirality, Eq.~\eqref{Chirality}, and by Eq.~\eqref{mave}.
For $n=16$ (Fig.~\ref{pippo1}b) we observe  
a structure more complex than that we find in the films with 
$n=12$ and $n=9$ (Figs.~\ref{pippo1}c and Fig.~\ref{pippo1}d
respectively). Indeed, as discussed in details in Ref.~\onlinecite{Cinti1},  
the 5$^{\textnormal{th}}$, and 6$^{\textnormal{th}}$ 
(and, by symmetry, the 11$^{\textnormal{th}}$ and 12$^{\textnormal{th}}$) 
planes lose their order at a lower temperature, $T_{C,16}(5,6,11,12)=117.8(2)$\,K, 
than the others, where
 $T_{C,16}(1\ldots4,7\ldots10,13\ldots16)=120.1(4)$\,K.

\begin{figure}
  \begin{center}
    \includegraphics[width=0.8\textwidth]{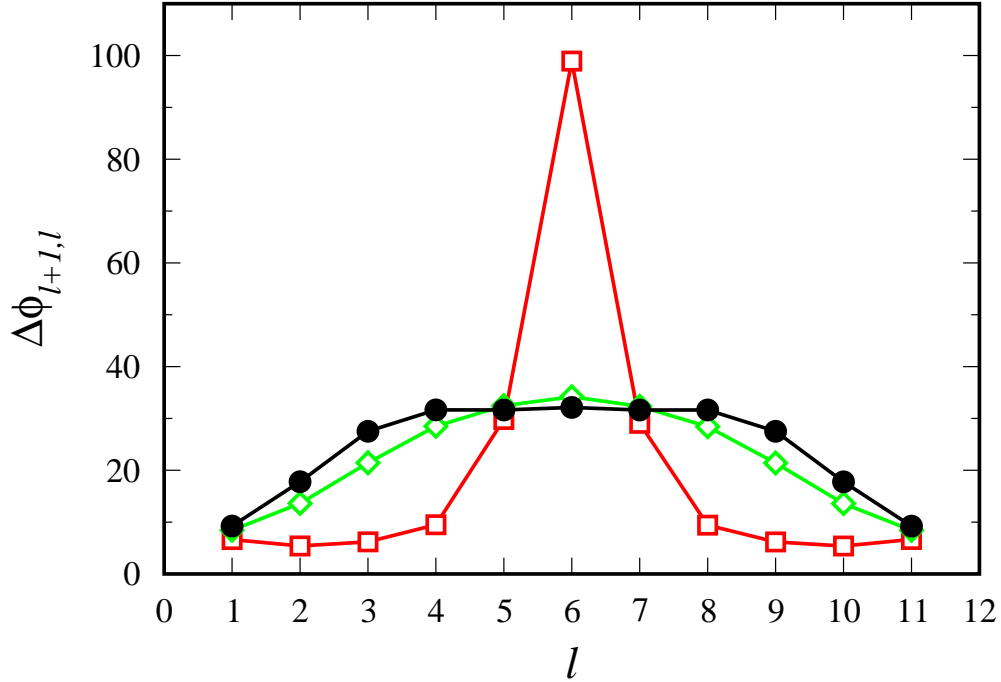}
  \end{center}
  \caption[]{(color online) Angle formed by the magnetization vectors 
    of NN planes, for $n=12$ and $L=48$ at different temperature: 
    $T=10$\,K (solid black circle), $T=110$\,K (open green diamond), 
    and $T=118$\,K (open red square)\cite{note2}. Error bars lye within point size.}
  \label{pippo2}
\end{figure}
In order to better understand the magnetic structure in
these temperature ranges, we examine the $T$-dependence of 
the magnetization rotation angle $\Delta\phi_l$. For the sake of clarity
in Fig.~\ref{pippo2} the system $n=12$ is again analyzed. 
When $T_{C,12}(6)<T<T_{C,12}(1\ldots5)$ (see, e.g., red square and line
in Fig.~\ref{pippo2}) the ordered layers distinctly display a block structure
where $\Delta\phi_l$ among the first (last) five planes is almost
zero, i.e. 
$\lesssim 10^{\circ}$; at the same time the
angle formed by the magnetization of the two blocks is about $180^\circ$\cite{note2}.
Only for $T<T_{C,12}(6)$ (black and green symbols and lines in
Fig.~\ref{pippo2}), the function $\Delta\phi_l$ displays the expected 
thin films helimagnetic behaviour discussed in Sec.\ref{lowt}
(for comparison, see Fig.~5a of Ref.~\onlinecite{Weschke04}, where 
the same quantity at $T=0$ is discussed within MFA).

We can graphically represent the block magnetization arrangement 
for $n=12$ as $\uparrow\uparrow\uparrow\uparrow\uparrow\circ\circ%
\downarrow\downarrow\downarrow\downarrow\downarrow$, 
where the circle represents the disordered planes, and the arrows the 
ordered ones. 
As above anticipated, the spin block phase is obtained down to $n=9$
(Fig.~\ref{pippo1}d), where we get
an arrangement $\uparrow\uparrow\uparrow\uparrow\circ%
\downarrow\downarrow\downarrow\downarrow$, and up to $n=16$
(Fig.~\ref{pippo1}b) where a much more 
intricate layout, i.e. $\uparrow\uparrow\uparrow\uparrow\circ\circ%
\downarrow\downarrow\downarrow\downarrow\circ\circ\uparrow\uparrow\uparrow\uparrow$,
is observed.
It is worthwhile to observe that the AFM alignment of 
consecutive ordered blocks reveals as the medium range 
alternating inter-layer exchange coupling give rise to an effective 
AFM interaction between blocks.

\begin{figure}
  \begin{center}
    \includegraphics[width=1\textwidth]{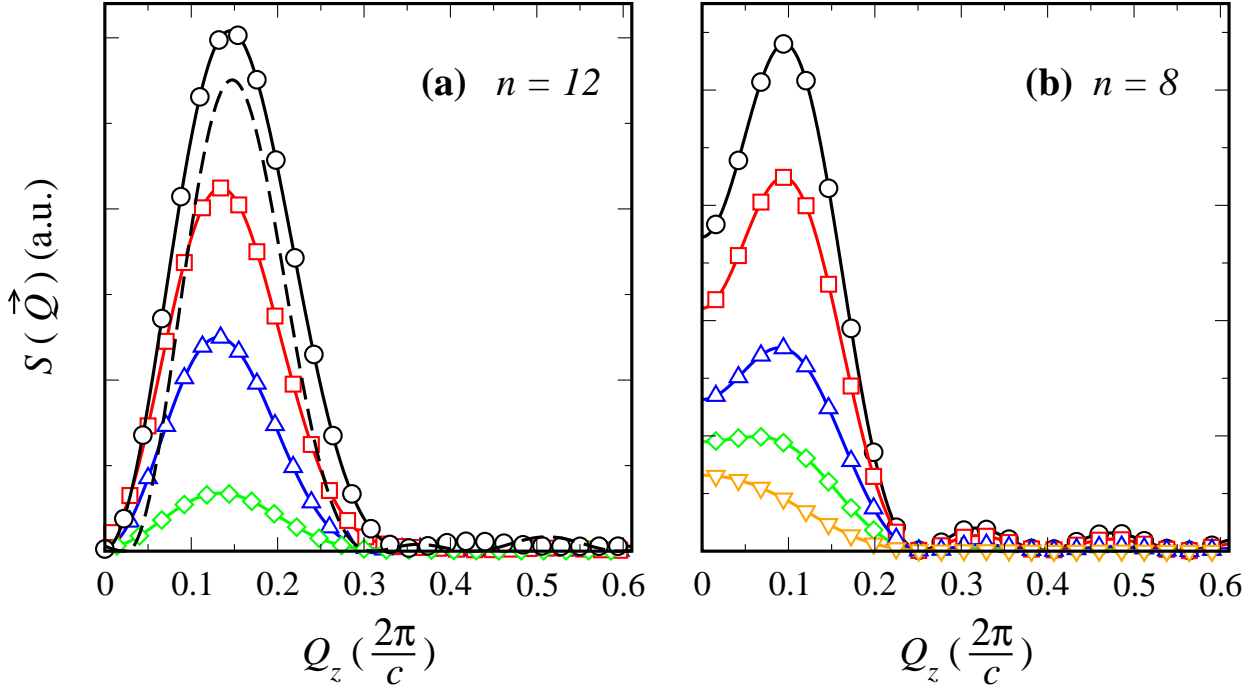}
  \end{center}
  \caption[]{(color online) Temperature evolution of $S(0,0,Q_z)$
    \textit{vs.} $Q_z$. \textbf{(a)}: thickness $n=12$ and $L=48$ at
    $T=10$\,K (black line and circle), 113\,K (red line and square), 116\,K
    (blue line and triangle), and 121\,K (green line and diamond). Dashed
    black line: $S(\vec{Q})$ of the saturated block structure
    $\uparrow\uparrow\uparrow\uparrow\uparrow\circ\circ\downarrow\downarrow\downarrow\downarrow\downarrow$. 
    Both the black curve ($T=10$\,K) and the dashed one have been divided
    by a factor equal to ten. \textbf{(b)}: thickness $n=8$ and $L=48$
    at $T=10$\,K (black line and circle), 50\,K (red line and square), 
    90\,K (blue line and up-triangle), 105\,K (green line and
    diamond), and 110\,K (orange line and down-triangle); all temperatures
    are lower than $T_{C,8}(1\ldots8)$.}
  \label{sqt}
\end{figure}

A further insight in these block phases, especially relevant
from an experimental point of view, is obtained by analyzing  
the behaviour of the structure factor close to $T_{C,n}$($l$).
In Fig.~\ref{sqt}a $S(\vec{Q})$ for $n=12$ in a wide temperature range
is plotted. In particular, one temperature value just below $T_{C,12}(6)$
($T=113$\,K, red line and square), one in the block phase region
$T_{C,12}(6)<T<T_{C,12}(1\ldots5)$ ($T=116$\,K, blue line and 
up-triangle), and one just above 
$T_{C,12}(1\ldots5)$ ($T=121$\,K, green line and diamonds)
have been chosen.
As already observed in Sec.~\ref{sec1}, the prominent broadening 
in the whole temperature range of the peaks displayed by $S(\vec{Q})$ 
is mainly a consequence of the intrinsic finite size nature of the films.
Excluding an obvious intensity reduction for increasing temperature,
one can immediately observe that both shape and peak
position are almost unchanged moving from  
block phases (e.g., $T=116$\,K data) to HM order 
(e.g., $T=113$\,K data).
A further confirmation of the last statement 
can be achieved by the comparison, again proposed in Fig.~\ref{sqt}a, between 
the Monte Carlo outcome for $S(\vec{Q})$ at $T=10$\,K (black line and circle) 
and the Fourier transform of the static block structure 
$\uparrow\uparrow\uparrow\uparrow\uparrow\circ%
\circ\downarrow\downarrow\downarrow\downarrow\downarrow$
with saturated magnetization for each ferromagnetic plane 
(black dashed line). As formerly observed for 
the ultra-thin film $n=16$ \cite{Cinti1}, even for $n=12$
the two plots have the same peak position and width;
moreover they have comparable intensities too.
We are thus led to conclude for the substantial impossibility 
to distinguish between block phases and helimagnetic order by 
looking at the structure factor, being it able to
give information about the global structure modulation only.

\begin{figure}
  \begin{center}
    \includegraphics[width=0.8\textwidth]{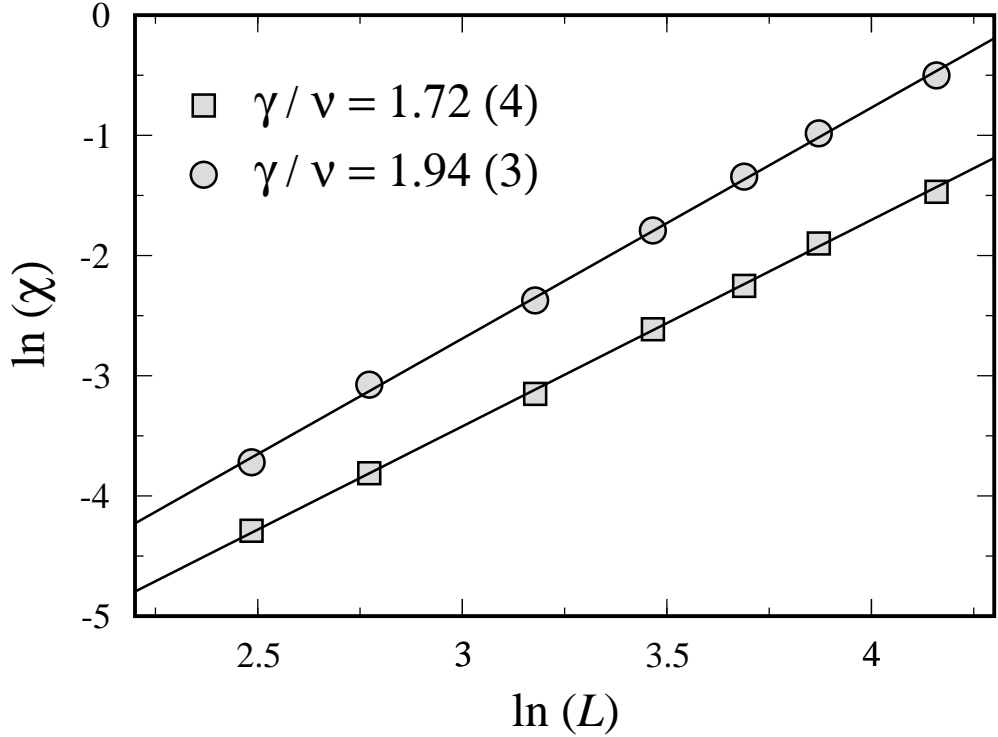}
  \end{center}
  \caption[]{(color online)
    Logarithm of the susceptibilities $\chi_{M_{5}}$ (square) and
    $\chi_{M}$ (circle)  \textit{versus} the logarithm of the lateral
    film dimensions $L$ (error bars included in the symbols), for
    $n=9$, at $T_{C,9}(5)$ and $T_{C,19}(1\ldots4,6\ldots9)$, respectively.
    }
  \label{3d-2d-scaling}
\end{figure}
% KT

As already observed in Ref.~\onlinecite{Cinti1} for $n=12$,
we find that whenever a blocked phase temperature range occurs,
the spins lying on disordered layers are seen to feel a
local magnetic field due to inter-layer interactions
much smaller than that acting on spins
on the ordered layers, so that they behave as
being effectively decoupled from the other ones and display
the characteristic features of a two-dimensional
magnet. The different effective dimensionality
of the critical behaviour of lowest-temperature ordering
layers from highest-temperature ordering ones, is illustrated
in Fig.~\ref{3d-2d-scaling}, where
an accurate finite-size scaling analysis of the layer magnetic
susceptibility $\chi_{M_{5}}$ and of the global susceptibility
$\chi_M$, at $T_{C,9}(5)$ at $T_{C,9}(1\ldots4,6\ldots9 )$,
respectively, is reported for $n=9$: making use of the usual
scaling relation at the critical temperature
$\chi \propto L^{\gamma / \nu}$, where $\gamma$ and $\nu$
are the critical exponents of the susceptibility and correlation length,
respectively, the value $\gamma/\nu=1.72$\,(4), is obtained
from the best fit of $\chi_{M_{5}}$ data, while
$\gamma/\nu=1.94$\,(3) is the result of the fit of
$\chi_M$. The former value is completely consistent with
the Kosterlitz-Thouless behaviour expected in an isolated
two-dimensional, easy-plane magnet, while the latter clearly indicates
a planar three-dimensional-like trend\cite{Janke93a,Janke93b}
for the system made by the planes $1\div4$ and $6\div9$. 

We now move to discuss the MC results obtained for $n=8$. From
Fig.~\ref{pippo1}e the lack of the ordered-disordered blocks
mixed structure at intermediate temperature is apparent, as a 
transition temperature common to all planes 
($T_{C,8}(1\ldots8)=110.6(2)$\,K) is found.
In Fig.~\ref{sqt}b $S(\vec{Q})$ \textit{vs.} $Q_z$ for different $T$ is shown: 
Close to $T_{C,8}(1\ldots8)$, 
we have $Q_z^{max}=0$, signalling the presence of a FM-like collinear 
structure. Subsequently, as the temperature lowers, 
and the FM spin arrangement opens towards a more stable 
fan structure, $S(\vec{Q})$ develops a peak at $Q_z^{max}=0.095$,
but still with a very strong contribution also at $Q_z=0$.

\begin{figure}
  \begin{center}
    \includegraphics[width=0.8\textwidth]{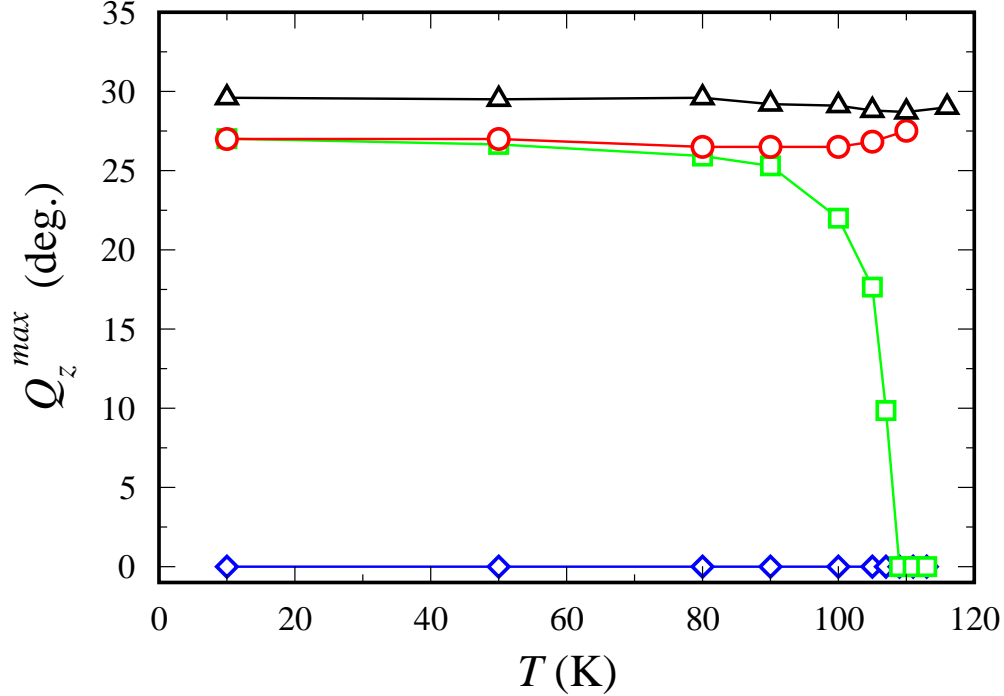}
  \end{center}
  \caption[]{(color online) $Q^{max}_z$ \textit{vs.} temperature for some
    thicknesses: $n=$\,16 (black line and triangle), 9 (red line and 
    circle), 8 (green line and square), and 7 (blue line and diamond). 
    For $n=8$ the stability of the fan phase with respect to the 
    collinear one is reached only below $T\approx 90$\,K (see text).}
  \label{evoluzioneq}
\end{figure}
The evolution of the structure-factor peak position with temperature
is better illustrated in Fig.~\ref{evoluzioneq}, where $Q_z^{max}$ \textit{vs.} $T$ 
is plotted for some significant values of film thickness. 
For $n=8$ the clear jump from a collinear structure to a fan-like one 
(reached at $T\approx 90$\,K) is observed:
This shows that the onset of order in every plane is by itself not 
necessarily enough to generate the fan structure observed at low temperature.
On the contrary, for thickness values close to the helical 
pitch and above (where $Q_z^{max}$ is essentially independent 
of temperature) the completion of planes ordering, with the 
transition of inner layers, marks also the onset of the overall 
helical/fan arrangement, while for small $n$ a ferromagnetic alignment
again stabilizes as soon as the layers simultaneously order.
Therefore, the peculiar behaviour of 
$Q_z^{max}(T)$ for $n=8$ can be reasonably attributed to its representing 
the borderline between helical/block ordered structures and
substantially ferromagnetic ones.

In view of the previous discussion, we can conclude that the 
existence of an intermediate temperature region, characterized by
the presence of spin block structures, in between the paramagnetic and 
the helical ones, seems to be a peculiar feature of non-collinear Ho
magnetic films with thickness close enough to the bulk Ho helical
pitch; we would like to emphasize that the allowance for at least six 
inter-layer interactions in the model Hamiltonian~\eqref{Hamiltonian} 
turns out to be essential in order to be able to observe such
behaviours \cite{note}.

%%%%%%%%%%%%%%%%%%%%%%%%%%%%%%%%%%%%%%%%%%%%%%%%%%%%%%%%%%%%%%%%%
%%%%%%%%%%%%%%%%%%%%%%%%%%%%%%%%%%%%%%%%%%%%%%%%%%%%%%%%%%%%%%%%%
%%%%%%%%%%%%%%%%%%%%%%%%%%%%%%%%%%%%%%%%%%%%%%%%%%%%%%%%%%%%%%%%%
 \subsection{Global Film Properties}\label{gfp}
In this Section we will analyse some macroscopic thermodynamic 
quantities of the film, and for clarity reasons the attention 
will be again focused mainly on $n=12$. 
We will show results pertinent to the magnetic specific heat, 
the chirality, Eq.~\eqref{Chirality},
and the average order parameter $M$ defined in Eq.~\eqref{mave}.
\begin{figure}
  \begin{center}
    \includegraphics[width=0.8\textwidth]{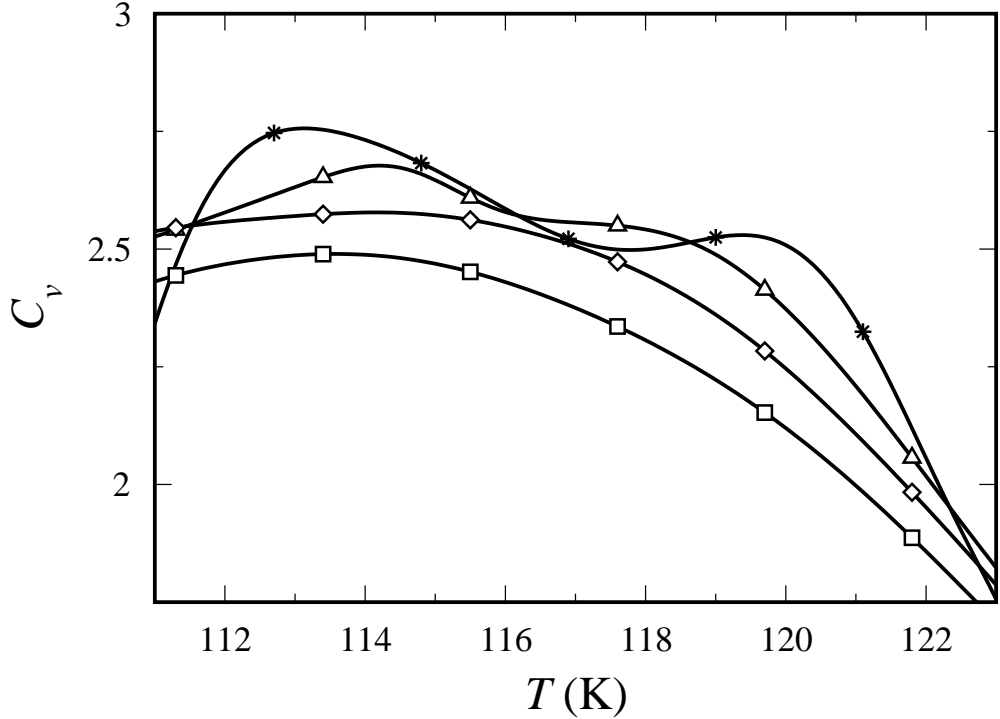}
  \end{center}
  \caption[]{Specific heat $C_v$ \textit{vs.} temperature for thickness 
    $n=12$ and lateral dimension
    $L=$\,48\,(star),\,32\,(triangle),\,24\,(diamond),\,16\,(square).}
  \label{cvfig}
\end{figure}

The first quantity we consider is the specific
heat. In Fig.~\ref{cvfig}, $C_v$ at
different $L$ is displayed. Its behaviour clearly suggests the presence
of two different phase transitions: In fact, two well separated maxima appear
in Fig.~\ref{cvfig} at $T=113.1$\,K and $T=119.4$\,K for $L=48$, 
i.e. close to $T_{C,12}(6)$ and $T_{C,12}(1\ldots5)$ respectively,
making the maxima clear footprints of the block phase regime.
Such feature could not be observed in Ref.~\onlinecite{Cinti1} for 
$n=16$ film: indeed, the thinner temperature range where the block
phase is present, joined with the broad character of the maxima for 
the finite-size samples investigated, made the two maxima coalesce 
and impossible to be resolved. Therefore,
differently from what happens for $n=12$, 
in that case the magnetic entropy seems completely released around 
$T_{C,16}(1\ldots4,7\ldots10,13\ldots16)$.

\begin{figure}
  \begin{center}
    \includegraphics[width=1\textwidth]{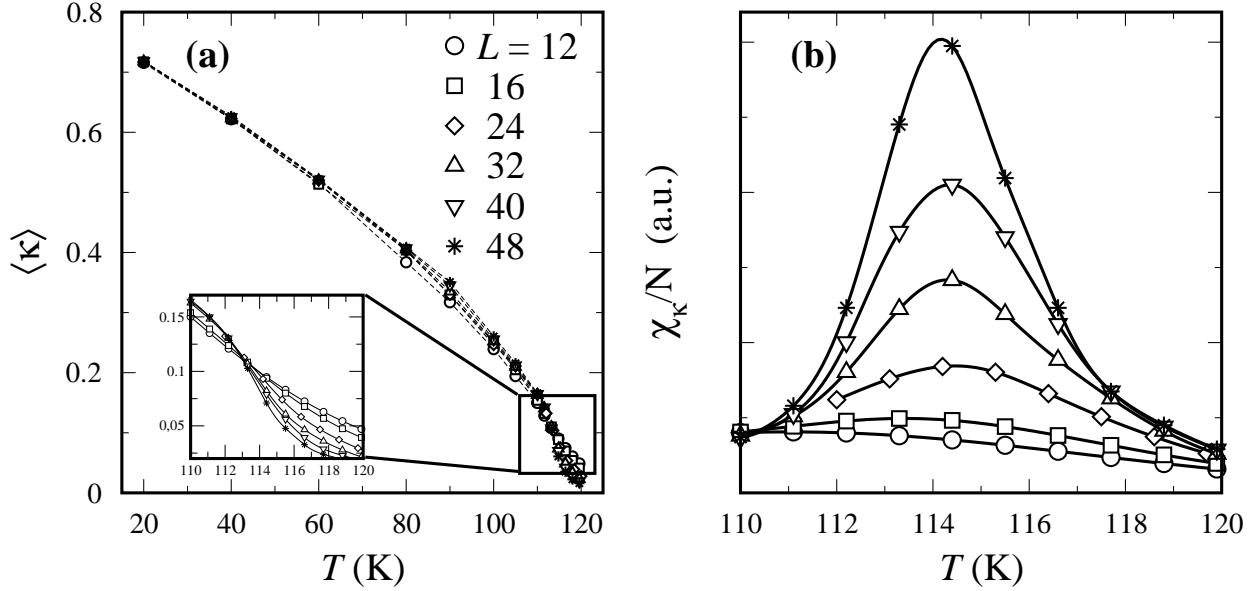}
  \end{center}
  \caption[]{\textbf{(a)} Chirality \textit{vs.} temperature at $n=12$.
    The normalization factor is fixed by the bulk factor 
    $\sin Q_z^{bulk}$ (see Eq.~\eqref{Chirality}); error bars are smaller than 
    point size. 
    The continuum lines in the inset are obtained by multiple-histogram 
    algorithm. 
    \textbf{(b)} $\chi_\kappa$ obtained by multiple-histogram algorithm. 
    The largest relative error is 0.5\% for $L=48$.}
  \label{kappa12op}
\end{figure}
The onset of a HM/Fan configuration along the perpendicular 
film direction can be probed by looking at a related quantity 
like the chirality, which is plotted in Fig.~\ref{kappa12op} 
together with its susceptibility.
$\chi_\kappa(T)$ does not show any anomaly 
in the proximity of the highest temperature maximum of $C_v$,
i.e. around 120\,K, while a clear peak appears at 
$T\simeq 114.2$\,K, becoming more and more sharp as $L$ increases.
\begin{figure}
  \begin{center}
    \includegraphics[width=1\textwidth]{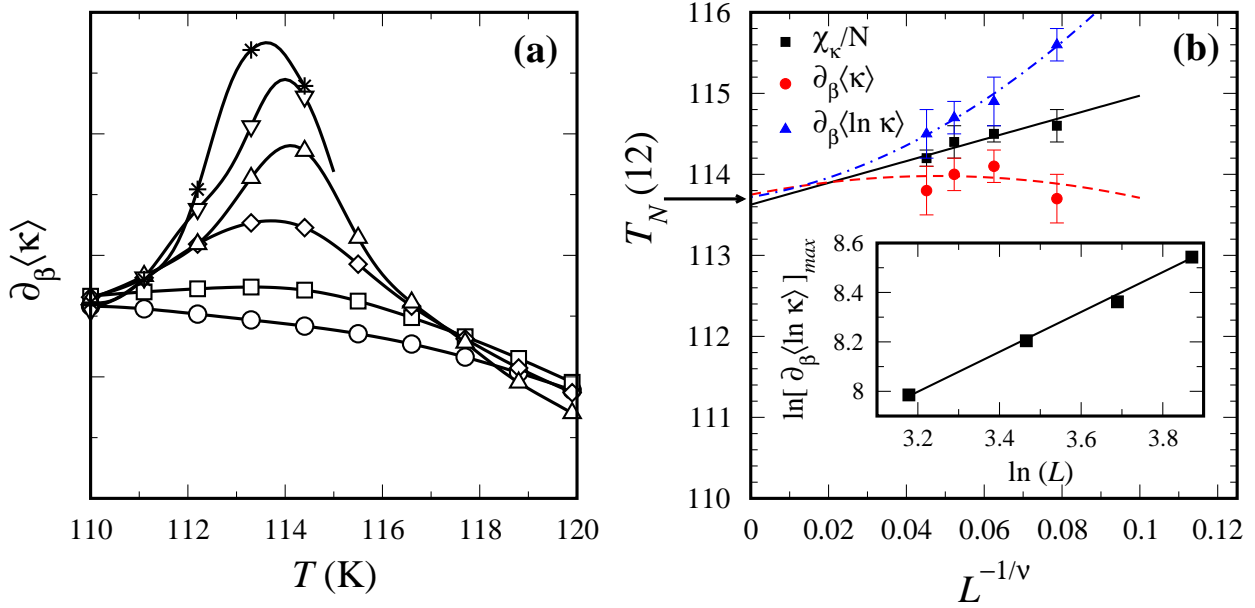}
  \end{center}
  \caption[]{\textbf{(a)}:  
    $\partial_\beta\langle\kappa\rangle$ \textit{vs.} temperature 
    for $L=$\,12,\,16,\,24,\,32,\,40,\,48 (symbols as in Fig.~\ref{pippo}).
    \textbf{(b)}: $T_N(12)$ plotted \textit{vs.} $L^{-1/\nu}$ 
    obtained by finite size scaling extrapolation of the three
    observables, with fitted $\nu$ value. Inset: plot of the maximum value of
    $\ln\left[\partial_\beta\langle\ln\kappa\rangle\right]$ 
    \textit{vs.} $\ln L$ together with its best fit function (see Eq.~\eqref{fitni}).}
  \label{der12}
\end{figure}
In order to estimate the transition temperature $T_N(12)$, 
a finite size scaling analysis of the quantities defined in the 
Eqs.~\eqref{susce}, \eqref{opprime}, and \eqref{lnopprime}, 
with  $\mathcal{O}=\kappa$, has been carried out.
In Fig.~\ref{der12}a $\partial_\beta\langle\kappa\rangle$ is 
reported to show the typical behaviour of such quantities, 
while Fig.~\ref{der12}b shows details of the fitting and extrapolation 
procedure.
We obtain $\nu$ using it as a free fit parameter in the equation \cite{Libbinder,bunker93}:
\begin{equation}
\label{fitni}
\partial_{\beta}\ln\kappa\left(t,L\right)=L^{1/\nu}\mathcal{X}\left(tL^{1/\nu}\right)
\end{equation} 
where $t=\frac{|T-T_N(12)|}{T_N(12)}$ and $\mathcal{X}$ is an
opportune scaling function. At the phase transition $T_N(12)$
(i.e. for $t=0$) we can consider the scaling relation~\eqref{fitni} as
$\left(\partial_\beta\ln\kappa\right)_{max}\propto L^{1/\nu}$.
Therefore, through an adequate fit, shown in the inset of 
Fig.~\ref{der12}b, we have obtained $\nu=0.79(2)$.
Using such value for $\nu$, we can estimate $T_{N}(12)$ from
Eq.~\eqref{tscaling} by looking at the
temperatures where $\langle\chi_\kappa\rangle$, 
$\partial_\beta\langle\kappa\rangle$, and 
$\partial_\beta\langle\ln\kappa\rangle$ acquire their maximum 
value and extrapolating them against
$L^{1/\nu}$ as shown in Fig.~\ref{der12}b.
The final result obtained is $T_{N}(12)=113.6(1)$\,K, a value 
definitely comparable with $T_{C,12}(6)=113.4(4)$\,K: we are thus lead to conclude
that the onset of a helical/fan order in the film is only possible when
all layers order, so that it is the last ordering spin layer, 
the 6$^{\textnormal{th}}$ one for $n=12$, that \textit{drives} the overall 
film transition to HM order.

An issue largely debated in literature \cite{Diep94,Loison00b,Diepold,Spanu}
concerns the order of the chiral transition: In our 
MCS, in the whole thickness range here analyzed, we
did not observe any double-peaked structure
in the equilibrium energy distribution at $T_N(n)$,
i.e., no explicit indication for a first order 
phase transition is given by our investigation.
Anyway a first order transition can not be completely excluded,
as suggested in Ref.~\onlinecite{Loison00b},
where the author reasonably observes that a firm evidence for 
a first order transition can be obtained 
only when the sample is much larger than the largest
correlation length \cite{Loison00b}.

\begin{figure}
  \begin{center}
    \includegraphics[width=0.8\textwidth]{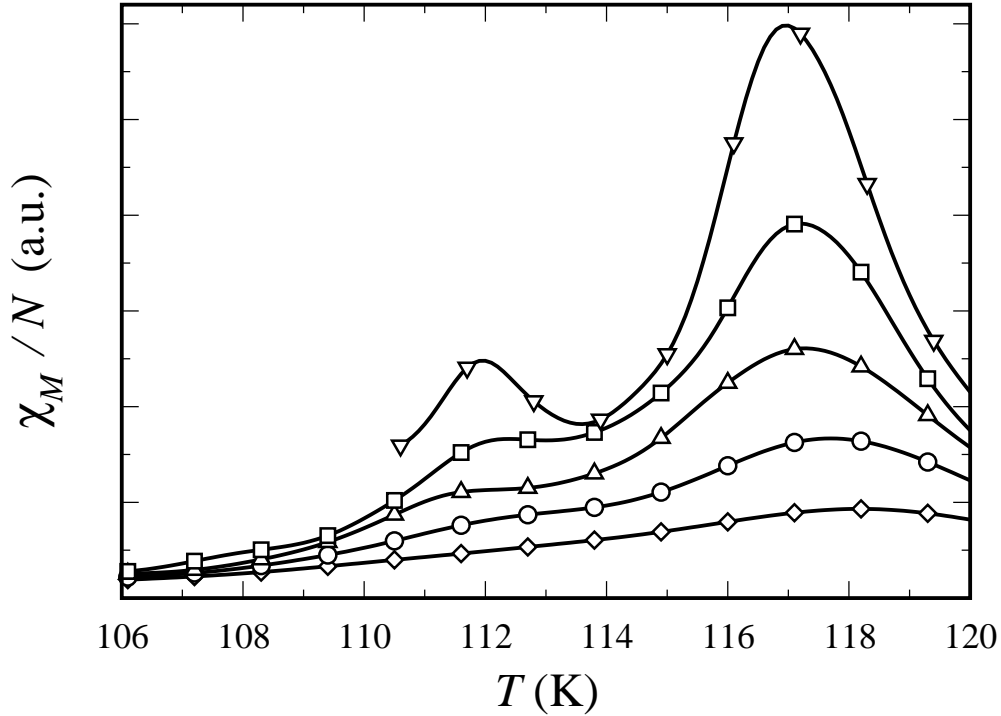}
  \end{center}
  \caption[]{Susceptibility $\chi_M$ at $n=9$ and 
    lateral dimensions $L=24$ (diamond), 32 (circle), 
    40 (up-triangle), 48 (square), and 60 (down-triangle)}
  \label{chim9}
\end{figure}
The average order parameter $M$ defined in Eq.~\eqref{mave} 
turns out to probe the physical properties of the system in a 
way more similar to what is done by the specific heat than 
by $\kappa$, bearing it signatures of the onset of both spin block
and HM/Fan phases, as it is apparent by looking at the related 
quantity $\chi_M$. As an example, $\chi_M$ at $n=9$ is reported 
in Fig.~\ref{chim9}:
two anomalies are present at $T\simeq 112.0$\,K and 
$T\simeq 117.0$\,K, i.e. at temperatures 
roughly corresponding to $T_{N}(9)=111.2(5)$\,K 
 and $T_{C,9}(1\ldots4)=115.9(4)$\,K. We may observe that 
while the qualitative behaviour of $\chi_M$ is similar to that
of the specific heat, the peaks in $\chi_M$ are sharper and display
the finite size scaling typical of a critical quantity, thus making
it a better probe to locate transition temperatures.

As the film thickness decreases, the chirality and its related observables display
a behaviour similar to that at $n=12$ up to $n=9$, despite a shift from HM to 
fan-like order (see Fig.~\ref{sqtot0}f); a largely different qualitative
behaviour is instead obtained for $n\leq8$, as already discussed in Sec.~\ref{planes}.

\section{Discussion and Conclusion}\label{disc} 

In this paper the magnetic properties of thin Ho films have been 
carefully investigated by extensive MCS,
assuming the model pertinent to bulk structure.
Regarding the magnetic order below $T_N(n)$,
it has been showed as, by decreasing the number $n$ 
of spin layers building the film, a progressive rearrangement of 
layer's magnetization from a helical to a fan-like structure
(i.e. $S^{xx}(\vec{Q})\ne S^{yy}(\vec{Q})$), and finally to an
essentially FM  order for $n\le 7$, is observed.
Moreover, for film thickness $n=8$ the structure factor analysis 
has clearly revealed that once a finite magnetization has been
established in every layer, an FM layer arrangement firstly appears which 
transforms to a fan-like configuration as the temperature is further reduced, 
as shown in Fig.~\ref{evoluzioneq}.

Above all that, the system presents very interesting
properties around the critical region when the film thickness
is comparable with the bulk holmium helical pitch, i.e. for
$9\le n \lesssim 16$.
A spin block phase regime is observed in a wide 
range of intermediate temperatures. In this window
some inner planes ($n=16$ has a more complex block
structure, as discussed in Ref.~\onlinecite{Cinti1}) are in a paramagnetic
configuration, while the other ones, close to the surfaces, appear 
to be in a quasi-FM ordered state
(for example, when $n=12$ we have obtained a spin block configuration 
where the magnetization rotate of an angle $\Delta \phi_l\sim 10^{\circ}$ 
when moving from one spin layer to a neighbouring one within the same
block, Fig.~\ref{pippo2}).
Everytime a spin block configuration appears, neighbouring
ordered blocks line up in an antiferromagnetic way.
What's more, also the study of macroscopic thermodynamic 
quantities, as the total energy, the order parameter defined 
in Eq.~\eqref{mplane} and their derivatives, confirms the presence 
of such large critical regions.

It is worth to remark that making use of all 
the six inter-layer coupling constants experimentally deduced by 
Bohr \textit{et al.} in Ref.~\onlinecite{Borh89},
it is seen that the competition among surface effects and
frustrated inter-layer interactions do not 
entail a simple adjustment of the surface planes only, but
the magnetic critical properties of the whole film 
are strongly modified as well.
Moreover, the results here presented, while 
confirming that most of the predictions of the MFA employed 
by Jensen and Bennemann in Ref.~\onlinecite{Jensen05} are 
qualitatively correct, also show unambiguously that the 
thermal fluctuations play an essential role, so that the
ability to include their effects is extremely important
in order to have a full comprehension of the block phase 
phenomenology in these films.

A detailed study of the chiraliy, Eq.~\eqref{Chirality},
has shown that $\kappa$ correctly describes
the establishing of a global helical/fan order
at $T_N(n)$ for $n\ge 9$, but such quantity does not result
critical in the temperature region
where the spin block phase structure appears.
Such behaviour of $\kappa$ is also observed in the 
borderline case $n=8$, where $\kappa$
does not present any anomaly at the single-layer ordering 
temperature $T_{C,8} (1\dots8)$.

Another important issue of non-collinear film structures is the 
impossibility to describe $T_N(n)$ \textit{vs.} $n$ through a  
well established scaling relation. In fact, as discussed in 
the Introduction, making use of the empirical relation~\eqref{filmhm}
one can easily only locate the thickness $n_0$ which 
signalizes the disappearance of the HM order.
From our results we can reasonably estimate $n_0=7\div9$: 
The partial disagreement with the experimental results 
\cite{Weschke04} ($n_0\approx 10$) 
could be a consequence both of defects and of the uncertainty
with which the thickness of the Ho film experimental samples 
is known \cite{comm}.

Finally, a possible comparison between the MCS results
and the experimental data \cite{Weschke04,Weschke05} would require
particular care, in order to avoid \textit{na\"\i ve} considerations.
Concerning the identification of the magnetic
order type, we have shown \cite{Cinti1} 
that the static structure factor alone is not able to 
distinguish between helical and spin block order.
At the same time we think that the characterization 
of such spin block phases in non-collinear magnetic 
thin films could be very useful for experimental future
works, being the coupling constants strongly
dependent both on the used deposition substrate 
and on the employed deposition technique \cite{comm,sper1}.

\begin{acknowledgments}
We would like to thank H.~Zabel for the fruitful discussions. 
\end{acknowledgments}


\begin{thebibliography}{99}

\bibitem{Fisher72} M.~E.~Fisher and M.~N.~Barber,
  Phys.~Rev.~Lett.~\textbf{28}, 1516 (1972).

\bibitem{Ritchie} D.~S.~Ritchie, and M.~E.~Fisher, Phys.~Rev.~B
  \textbf{7}, 480 (1973).

\bibitem{Barber} M.~N.~Barber, in \textit{Phase Transition and
    Critical Phenomena}, edited by C.~Domb and J.~Lebowitz (Academic,
  New York, 1983), Vol.8, Chap.~2.

\bibitem{Henkel99}  M.~Henkel, \textit{Conformal Invariance and
    Critical Phenomena} (Springer, 1999).

\bibitem{Henkel98} M.~Henkel, \textit{et al.},
  Phys.~Rev.~Lett.~\textbf{80}, 4783 (1998), and references therein.

\bibitem{zhang01} R.~Zhang and R.~F.~Willis,
  Phys.~Rev.~Lett.~\textbf{86}, 2665 (2001), and references therein.

\bibitem{Jensen91}  P.~J.~Jensen, and A.~R.~Mackintosh, \textit{Rare
    Earth Magnetism (Structure and Excitations)} (Clarendon Press,
  Oxford, 1991).

\bibitem{mf1} S.~W.~Cheong and M.~Mostovoy, Nature Materials (London)
  \textbf{6}, 13 (2007).

\bibitem{mf2} S.~Ishiwata \textit{et al.}, Science \textbf{319}, 1643
  (2008).

\bibitem{mf3} M.~Fiebig, J.~Phys.~D \textbf{38}, R123 (2005).

\bibitem{MnSi} C.~Pfleiderer \textit{et al.}, Nature (London) \textbf{414}, 427 (2001),
Nature (London) \textbf{427}, 227 (2004).

\bibitem{FeGe} P. Pedrazzini \textit{et al.}, Phys. Rev. Lett. \textbf{98}, 047204 (2007).

\bibitem{Jensen05} P.~J.~Jensen, and K.~H.~Bennemann, Surface Science
  Reports \textbf{61}, 129 (2006).

\bibitem{Bland} \textit{Ultrathin Magnetic Structures}, edited by
  J.~A.~C.~Bland, and B.~Heinrich (Springer, New York, 1994).

%\bibitem{qbit} S.~Bertaina, Nature Nanotechnology  \textbf{2}, 39 (2007).

\bibitem{Weschke04} E.~Weschke \textit{et al.},
  Phys.~Rev.~Lett.~\textbf{93}, 157204 (2004).

\bibitem{Weschke05} E.~Weschke \textit{et al.}, Physica B
  \textbf{357}, 16 (2005).

\bibitem{schss00} C.~Sch\"u\ss{}ler-Langeheine \textit{et al.},
  Journal of Electron Spectroscopy and Related Phenomena
  \textbf{114-116}, 953 (2001); Phys.~Rev.~Lett.~\textbf{84}, 5624
  (2000).

\bibitem{Fullerton95} E.~E.~Fullerton \textit{et al.},
  Phys.~Rev.~Lett.~\textbf{75}, 330 (1995).

\bibitem{Cinti1} F.~Cinti, A.~Cuccoli, and A.~Rettori, Phys.~Rew.~B
  \textbf{78}, 020402(R) 2008.

\bibitem{Borh89} J.~Bohr \textit{et al.}, Physica B \textbf{159}, 93
  (1989).

\bibitem{Gibbs85} D.~Gibbs \textit{et al.}, Phys.~Rev.~Lett.~\textbf{55}, 234 (1985).

\bibitem{Jensen87} C.~C.~Larsen, J.~Jensen, and A.~R.~Mackintosh,
  Phys.~Rev.~Lett.~\textbf{59}, 712 (1987).

\bibitem{Coqblin77} B.~Coqblin, \textit{The Electronic Structure of
    Rare-Earth metal and Alloys}  (Academic press, 1977).

\bibitem{comm} H.~Zabel (private communicatio).

\bibitem{moriya60} T.~Moriya, Phys. Rev. \textbf{120}, 91 (1960).

\bibitem{Dagotto06} I. A. Sergienko, and E. Dagotto, Phys. Rev. B
  \textbf{73}, 094434 (2006).

\bibitem{Mostovoy06} M.~Mostovoy, Phys. Rev. Lett. \textbf{96}, 067601 (2006).

\bibitem{super} S. V. Grigoriev, Yu. O. Chetverikov, D. Lott, and
  A. Schreyer, Phys. Rev. Lett. \textbf{100}, 197203 (2008).

\bibitem{met53} N.~Metropolis \textit{et al.}, J.~Chem.~Phys.
  \textbf{21}, 1087 (1953).

\bibitem{over87} F.~R.~Brown, and T.~J.~Woch, Phys.~Rev.~Lett.,
  \textbf{58}, 2394 (1987).

\bibitem{bootstrap} M.~E.~J.~Newman, and G.~T.~Barkema, \textit{Monte
    Carlo Methods in Statistical Physics} (Clarendon Press, Oxford 1999).

\bibitem{Ferrenberg88} A.~M.~Ferrenberg, and R.~H.~Swendsen,
  Phys.~Rev.~Lett.~\textbf{61}, 2635 (1988), Phys.~Rev.~Lett.~\textbf{63}, 1195
  (1989).

\bibitem{PRElandau} A.~M.~Ferrenberg, D.~P.~Landau, and
  R.~H.~Swendsen, Phys.~Rev.~E \textbf{51}, 5092 (1995).

\bibitem{bootstrap1} M.~E.~J.~Newman, and R.~G.~Palmer,
  J.~Stat.~Phys.~\textbf{97}, 1011 (1999).

\bibitem{Kawamura} H.~Kawamura, J.~Phys.: Cond.~Matt.~\textbf{10},
  4707 (1998).

\bibitem{KTt} See e.g. D.~R.~Nelson, in \textit{Phase Transion and Critical
    Phenomena}, edited by C.~Domb and J.~Lebowitz (Academic,
  New York, 1983), Vol.7, Chap.~1, and references therein.

\bibitem{Janke93a} C.~Holm, and W.~Janke, Phys.~Rev.~B \textbf{48},
  936 (1993).

\bibitem{Janke93b} W.~Janke, and K.~Nather, Phys.~Rev.~B \textbf{48},
  15807 (1993).

\bibitem{wang89} K.~Binder, and J.~S.~Wang, J.~Stat.~Phys.~\textbf{55},
  89 (1989).

\bibitem{bunker93}A.~Bunker, B.~D.~Gaulin, and C.~Kallin, Phys.~Rev.~B
  \textbf{48}, 15861 (1993),  Phys.~Rev.~B \textbf{52}, 1415 (1995).

\bibitem{Diep94} \textit{Magnetic Systems with Competing
    Interactions}, edited by H.~T.~Diep (World Scientific, 1994).

\bibitem{Cinti2} F.~Cinti \textit{et al.},
  Phys.~Rev.~Lett. \textbf{100}, 057203 (2008), and references therein.

\bibitem{Loison00a} D.~Loison, and P.~Simon, Phys.~Rev.~B \textbf{61},
  6114 (2000).

\bibitem{Loison00b} D.~Loison, Physica A \textbf{275}, 207 (2000).

\bibitem{Libbinder} D.~P.~Landau, and K.~Binder,  \textit{A Guide to
    Monte Carlo Simulation in Statistical Physics} (Cambridge
  University Press, Cambridge, 2000).

\bibitem{binderu4} K.~Binder, Z.~Phys.~B \textbf{43}, 119 (1981).

\bibitem{binderprl}K.~Binder, Phys.~Rev.~Lett.~\textbf{47}, 693
  (1981).

\bibitem{note2} Bearing in mind the Binder cumulant analysis of 
Fig.~{\protect{\ref{pippo}}},
$\Delta\phi_{5,6,7}$ at $T=118$~K
are really meaningless, because the 6$^{\textnormal{th}}$ and
7$^{\textnormal{th}}$ layer still are in a substantially paramagnetic state; the relative
points in Fig.~{\protect{\ref{pippo2}}} (red square and line) are 
reported for continuity, and only the sum
$\Delta\phi_5+\Delta\phi_6+\Delta\phi_7=\phi_8-\phi_5$ is 
to be considered physically sound.

\bibitem{note} In a model Hamiltonian where two 
  inter-layer coupling constant only
  ($J_1$ and $J_2$ along the $c$-directions) are taken into account,
  a substantially different behaviour is observed: this topic will be 
  more detailly addressed and discussed in a forthcoming paper. 

\bibitem{Diepold} H.~T.~Diep, Phys.~Rev.~B \textbf{39}, 397 (1989).

\bibitem{Spanu} R.~Quartu, and H.~T.~Diep, Journal of Magnetism and
  Magnetic Materials \textbf{182}, 38 (1998). 

%\bibitem{Landau91} P.~Peczak, A.~M.~Ferrenberg, and D.~P.~Landau,
%  Phys.~Rev.~B \textbf{43}, 6087 (1991).

%\bibitem{Jensen90} P.~J.~Jensen, and A.~R.~Mackintosh,
%  Phys.~Rev.~Lett.~\textbf{64}, 2699 (1990).

\bibitem{sper1} J.~Kwo, in \textit{Thin Film Techniques for Low-Dimensional
  Structure}, edited by R.~F.~C.~Farrow, S.~S.~P.~Parkin, P.~J.~Dobson,
  N.~H.~Neaves, and A.~S.~Arrott (Plenum, New York, 1988) NATO ASI Ser.~B Vol.~13.


\end{thebibliography}
\end{document}